# Above Room-Temperature Phase Transition in Helicoidal 2D Halide Perovskite Enables Pyro-Phototronic Control

Zinnia Mallick[a], Rajashi Haldar[b,†], Sudip Naskar[a,†], Bapan Jana[b], Maheswaran Shanmugam[b,*], Shanker Ram[c,*], and Dipankar Mandal[a,*]

[a] *Quantum Materials and Devices Unit, Institute of Nano Science and Technology, Knowledge City, Sector 81, Mohali 140306, Punjab, India*

[b] *Department of Chemistry, Indian Institute of Technology Bombay, Mumbai, Maharashtra 400076, India*

[c] *Materials Science Centre, Indian Institute of Technology, Kharagpur, 721302 India*

[†] *Contributed equally*

[*] Email: eswar@chem.iitb.ac.in, sram@matsc.iitkgp.ac.in, dmandal@inst.ac.in

## ABSTRACT

Bridging pyro-phototronic and ferroelectric properties in a single material not only gives rise to exotic physical phenomena but also provides strategy to build-up next-generation multi-functional energy harvesting devices. The layered halide perovskites are emerging class of synergetic 2D-materials of exceptional light-induced functionalities. Here, we report synthesis of a lead-free perovskite (3-fluorobenzylamine)$_2$CuCl$_4$ of partially fluorinated aromatic rings. Molecular chains are chiral in helicoids so to render flexibility, energy-transfer, and freedom to tailor tunable pyro-phototronics on light illumination. The fluorinated framework promotes ferroelectric-to-paraelectric transition point ($T_C$), as high as ~ 412 K of wide range of workably. A giant pyro-photronic response is distinctly evident even upon UV-visible light–illuminations. A significantly high pyro-photronic current of 60 nA was achieved under an illumination of $\lambda_{ex}$~ 365 nm (27 mW) in a self-powered configuration. Consequently, an intricate coupling of the spontaneous polarization to the optical properties is visualized from the piezo response force microscopy (PFM) responses. In particular, the polar domains get diminished reversibly in on-off steps of the light irradiations. It indicates, the light provides another degree of the freedom to control the features for the applications of optoelectronic devices, optical memories, photo/thermo-chromic systems, and energy-harvesters.

## 1. INTRODUCTION

The ferroelectrics – a specific class of energy-materials exhibit switchable spontaneous electronic polarization in response to applied electric fields.[1-4] Nowadays, they have become key components of modern electronics of memory devices, capacitors, micro- actuators, biosensors, and many others.[1-4] A premier application outlet of synergetic ferroelectrics is the pyroelectric effect that leads to transform thermal-energy into electric-energy and vice versa.[3,5] It paves great promises in self-powered energy harvesters, pyro-photodetectors, medical diagnosis, real-time health monitors, flexible electronics, and beyond.[6–10] Ferroelectricity on being integrated to other synergetic properties in a material, such as magnetism, pyroelectricity, and optical properties, opens great potentials of multi-functionalities for many applications. Such phenomena give the freedom to design smart devices for next generation photovoltaics, nanosensing, and optoelectronic devices.[11-17] Ferroelectricity combined to optical and semiconducting properties inspires pyro-phototronic effect and bulk-photovoltaic effect (BPVE), which are required to develop self-powered photodetectors, photothermal sensors, and photovoltaic devices.[17-19] The pyro-phototronic effect paves light-induced pyroelectricity coupled to excitons in polar pyroelectrics.[19,20] It modulates processes of generation of excitons, charge separation, charge-recombination, and motions of the ferroelectric domains. For example, it is shown to reduce the response time of ZnO-based self-powered photodetectors by orders. Leveraging pyro-phototronics, many self-powered photodetectors of polar pyroelectrics, such as ZnO,[20,21] CdS,[22] $Ga_2O_3$,[19] Cu(In,Ga)$Se_2$,[19,23] have demonstrated fast response times and relatively high photoresponsivity. However, many high-performance optoelectronic semiconductors (e.g., Si, GaAs, GaN, etc.) are non-polar that limit their applications in self-powdered pyro-phototronic devices.

In this regard, molecular ferroelectrics have gained wide-spread attentions for their unique advantages of mechanical flexibility, lightweight, easy processability, and low acoustic impedance. That are being treated as potential supplementary of the traditional ceramic ferroelectrics such as $BaTiO_3$, $LiNbO_3$, PZTs, etc.[24,25] The ceramics are fragile and suffer from high band-gap ($E_g$), large internal resistance, and poor carrier transport. In this context, the layered organic-inorganic halide perovskites (OIHPs) of quantum-well structure and tunable $E_g$ are emerging as a superior material choice.[11,26] A general chemical formula $R_2A_{n-1}B_nX_{3n+1}$ describes the 2-dimensional (2D) and quasi-2D layered perovskites, where R is a large organic cation (aliphatic or aromatic ammonium) that separates the inorganic octahedral unit sheets ($B_nX_{3n+1}$).[14,27] Usually, A is a smaller organic radical, B is a metal ion, X is halide (Cl, Br, or I), and the index 'n' refers to the number of octahedral sheets are co-bonded in a network. For a strictly 2D OIHP (n = 1), it assumes a chemical formula of $R_2BX_4$.[27] Unlike the 3D perovskites, the 2D counterpart with alternating organic and inorganic layers breaks the strict limit of Goldschmidt tolerance factor and offers greater structural flexibility for the materials design of tailored $E_g$ value. The $BX_4$ octahedral layers offer band-gap engineering, high carrier mobility, large optical absorption, and functionalities. While the dynamic organic unit offers a large degree of freedom for molecular motions, which lead to symmetry breaking and enhance the exotic properties. Thus, 2D OIHPs present a unique opportunity to engineer novel halide perovskites with sustainable optoelectronic and ferroelectric properties.[12,14,27-29]

Naming a few, (benzylammonium)$_2$PbCl$_4$,[30] (cyclohexaylammonium)$_2$PbBr$_4$,[31] and (4,4-difluorocyclohexylammonium)$_2$PbI$_4$.[32] are designed in advancing the model optical and ferroelectric properties. Oftentimes, the OIHPs suffer from low ferroelectric–to-paraelectric transition temperature ($T_C$), and a strategy of a partial fluorination of the aromatic rings is shown to promote the $T_C$ value. For instance, the $T_C$ of PFBA$_2$PbBr$_4$ (440 K)[33] (PFBA = bis(perfluorobenzylammonium) and 4-FABCHCdCl$_3$ (419 K) (4-FABCH = 4-fluoro-1-azabicyclo2.2.1heptane) [34] are considerably enhanced from the non-fluorinated skeletons BA$_2$PbBr$_4$ (BA= n-butylammonium) (405 K) and ABCHCdCl$_3$ (190 K) (ABCH = 1-azabicyclo2.2.1heptane). In quite few instances, ferroelectricity is inspired in fluorinated perovskites 4,4-DFHHA$_2$PbI$_4$, (4,4-DFHHA = 4,4-difluorohexahydroazepinium) [35] and (TMFM)FeBr$_4$ (trimethylfluoromethylammonium)[36] compared to their otherwise non-fluorinated skeletons. Apparently, highly negative $2p^5$-F polarity tunes a highly polar phase on negative $BX_4$ poles at the opposite molecular sides of lowered symmetry. Thus, it is a model molecular design of a deeply polar ferroelectric phase.

In this article, we report a fluorinated 2D layered $Cu^{2+}$-$3d^9$ ($^2D_{5/2}$ spin state at net spin S = ½) perovskite (3-fluorobenzylamonium)$_2$CuCl$_4$ (3-FBA)$_2$CuCl$_4$), which adopts a non-centrosymmetric orthorhombic $Pca2_1$ crystal structure. A substantially high $T_C \sim 412$ K is tailored in due effects of the F atoms. Uniquely, the 2D-layers (width w $\rightarrow$ 8.3 nm) and molecular chains are chiral (pave transfer of light-energy into electro-mechanical energy) of a high-entropy phase (helicoids). Polarized domain dynamics studied with piezoresponse force microscopy (PFM) in dark and in light-illumination confer the ferroelectric phase**.** A systematic study of the light-regulated spontaneous polarization, pyroelectricity and self-powered photodetection reveals that the domains are built-in through pyro-phototronic effect. A ferromagnetic order prevails at room temperature.

## 2. RESULTS AND DISCUSSION

### 2.1 Surface orders in $(3\text{-FBA})_2CuCl_4$ layers

Thin plates of $(3\text{-FBA})_2$-$CuCl_4$ crystals (yellowish colour) were grown by slowly evaporating a precursor solution of 3-FBA and $CuCl_2{\cdot}2H_2O$ at room temperature. The synthesis procedure is described in the experiential section (given in the Supporting Information). Stacking of the atoms along *a*, b and c-crystallographic axes are briefed in Figure S1a,b,c in an orthorhombic structure. The XRD patterns measured at low temperature (such as 150 K) and room temperature do not show much change, except a 1.50% thermal expansion of the unit cell volume $V_c = 1.7988$ nm$^3$ (at 300 K) on heating 150 K to 300 K. Other structural parameters are given in Table S1. Hence, all other studies are carried out on a polar $(3\text{-FBA})_2CuCl_4$ of an orthorhombic crystal structure of $Pca2_1$ space group (point group mm2) at room temperature, which is very similar to the ferroelectric polar structure of a well-known ferroelectric $HfO_2$.[37] As projected in schematics in Figures 1a, b, c, the $(3\text{-FBA})_2$-$CuCl_4$ crystal structure consists of $CuCl_4^{2-}$ single-layers of corner-sharing $(CuCl_6)^{4-}$ octahedra bonding to 3-FBA$^+$ bilayers in the form of a 2D organic-inorganic $A_2BX_4$ perovskite structure (Ruddlesden–Popper phase), where the 3-FBA$^+$, $Cu^{2+}$ and $Cl^-$ ions correspond to the A, B and X sites, respectively. Figure 1c clearly shows that in the crystal packing, the 2D layers of both $CuCl_4^{2-}$ and 3-FBA$^+$ are in tilted positions with other layers, along with the different positioning of F atoms in the 3-FBA$^+$ bilayers. So, no inversion symmetry lasts in a $(3\text{-FBA})_2CuCl_4$ hybrid structure.

The inner $Cu^{2+}$ in $(3\text{-FBA})_2CuCl_4$ shown in Figure 1d experience the Jahn-Teller distortion, a common scenario for 2D-layered halide perovskites,[34-36] wherein two Cu–Cl bond lengths (2.929(1) and 2.986(1) Å) are elongated over the other four ones (2.293(1)–2.313(1) Å) (Jahn-Teller-out distortion, Figure S2a,b). The selected bond angles are given in Table S2. The $CuCl_6^{4-}$ octagons are bridged in 2D-layers through strong intra- and intermolecular N−H···Cl hydrogen bonds. The hydrogen bonding (create local charge) is effectively stronger on one side (H---Cl ~ 3.331(1) Å) with one of the 3-FBA moiety, as compared to the other one (H---Cl ~ 3.382(1) Å) from the other 3-FBA moiety on an opposite side of the molecular plane. The organic bilayers share weak inter- and intramolecular C-H···F bonding (H---F ~ 3.382(1) Å) (Figures S3c,d and Table S3). Such robust hydrogen-bonding not only reforms a non-centrosymmetric $(3\text{-FBA})_2CuCl_4$ network of a hierarchical structure**,** but also stiffens its thermal stability, $T_C$ point, and in turn boosts coupled pyro-phototronic and ferroelectric properties (described further) in an inorganic-organic framework. The $CuCl_6^{4-}$ chains (Figure 1e), as on bonded in 2D layers (helicoids), keep the $3d^9$-$Cu^{2+}$ spins (S = 1/2 spin/atom) coupled in a ferromagnet (FM) state. As discussed later with the XPS/Raman bands, part of $Cu^{2+}$ is thus reduced to $Cu^{+}$–$3d^{10}$ (S = 0, diamagnetic, DM), which can bond over a FM-$Cu^{2+}$ layer. Here, part of $CuCl_6^{4-}$ octagons is swiped to $CuCl_4^{3-}$ tetragons. An energy-level diagram in Figure 1f displays the $Cu^{2+}$ ground electronic state multiplet $^2D_J$ (J = 3/2, 5/2) is displaced over the $Cu^+$ state $^1D_2$, which are co-bridged in a network of switchable $^1D_2 \leftrightarrow {}^2D_J$ spin-states.

The field-emission scanning electron microscopy (FESEM) and atomic force microscopy (AFM) images in Figures 2a, b illustrate $(3\text{-FBA})_2$-$CuCl_4$ is ordered in layers (thickness t = 100−500 nm) one over others at L = 20−50 μm length scale (w = 10−20 nm) in shapes of thin laminates. They are exfoliated (Figure 2b) of well-smooth surfaces (RMS surface roughness of ~300 pm) at molecular scale. The surface topology is found to be devoid of any discontinuities. A closer view shows t = 5-10 nm thin layers (Figure 2c) of the transmission electron microscopy (TEM) images. A topology of region A (Figure 2c) zoomed in Figure 2d presents how the molecular chains are chiral in helicoids (w ~ 8.1 nm) at a spacing of 8.6 nm (less-darkish) in form of a molecular grating. A closer view of helicoids (added in the inset) insights tiny dots (w = 0.5 − 1.0 nm) are displaced zigzag like toroidals. Single $(3\text{-FBA})_2$-$CuCl_4$ moieties look like small dots and that are bonded in chains (chiral), which are coiled in helicoids (Figure 1e) at a coarser scale. This is highlighted with magnified TEM images (Figure 2e) of model strings (added in the inset). The HRTEM (high-resolution TEM) images in Figures 2f, g, h display packing of atoms at $d_{200}$ = 0.3550 nm, $d_{020}$ = 0.3750 nm, and $d_{008}$ = 0.4120 nm interplanar spacing are crumpled in polar toroidals. Such toroidal polar topology is observed in highly strained ferroelectric polymers of anticoupled chiral domains.[38] Here, $d_{200}$ and $d_{020}$ are reduced in a compressive strain $\sigma_s \rightarrow 2.69\%$ from the bulk values of 0.3648 nm and 0.3787 nm, respectively, while the $d_{008}$ is 1.25% stretched from the bulk state (Table S1). An asymmetric axial strain distorts the lattice to no longer retain centrosymmetry, a prerequisite to induce the electric polarization. Figure 2i projects model pathways of molecular chains (chiral) are wrapped one to others as 1D helicoids (as observed in Figure 2d), which edge-on bond as 2D layers.

The effects of $(3\text{-FBA})_2CuCl_4$ is exfoliated in thin layers result in due changes in the Raman bands of three major groups; (i) 30-300 $cm^{-1}$, (ii) 400-1750 $cm^{-1}$, and (iii) 2850-3250 $cm^{-1}$. The low frequency bands (Figure 3a) of $(CuCl_6)^{4-}$ moieties exhibit an order of larger intensities than the higher frequency bands of the organic network. Analogous to earlier reports on similar OIHPs,[39-42] out of six bands in region (i) at 60, 78, 102, 177, 248 and 276 $cm^{-1}$, the bands of 177 and 248 $cm^{-1}$ describe the equatorial and axial in-plane oscillations of Cu-Cl bonds, respectively, while the 276 $cm^{-1}$ band describes asymmetric Cu-Cl bond stretching in the $(CuCl_6)^{4-}$ moieties. The 102 $cm^{-1}$ band presents torsion of '$\bullet NH_3$' heads bond to the aromatic rings. A similar band is also shown in $(CH_3NH_3)_2$-$CuCl_4$ perovskite.[42] The last two bands of 60 $cm^{-1}$ and 78 $cm^{-1}$ are likely $CuCl_4$ liberations.[41,42] Alternatively, the $Cu^{2+}$-$3d^9$: $^2D_{5/2}$ state can split in '2J+1 = 5' Stark levels (crystal field splitting), and can share similar transitions (as the spin-waves). As included in the inset (Figure 3a), chirality[43] of polar $(3\text{-FBA})_2CuCl_4$ chains is likely to inspire the low energy spin-waves of spintronics. The fluorobenzyl units share (Figure 3b) two strong bands at 731 $cm^{-1}$ (C-F stretching) and 1002 $cm^{-1}$ (C-C ring breathing),[44] while the '$\bullet NH_3$' units share bit weaker bands of symmetric (1453 $cm^{-1}$) and asymmetric (1561 $cm^{-1}$) N-H bending,[45,46] mixed with the C=C ring stretching. Four C-H stretching modes at the C-C rings are shown at 2888, 3002,3062 and 3086 $cm^{-1}$ (Figure 3c) with a symmetric (2968 $cm^{-1}$) and an asymmetric (3145 $cm^{-1}$) N-H stretching at $\bullet NH_3$ units. The N-H stretching bands (rather intense) are broadened in hydrogen and van der Waals (vdWs) bonds to the $CuCl_4$ moieties in the inorganic-organic networks. Also C-H-stretching of 2888 $cm^{-1}$ is softened from 3000-3100 $cm^{-1}$ in benzene skeleton,[44] in due energy-loss to hydrogen bonds.

Thin $(3\text{-FBA})_2CuCl_4$ layers ($t \geq 225$ nm) show a marked effect of mode softening (by $\leq$ 15 $cm^{-1}$) and surface enhanced intensity of liberations (Figure 3d), while the normal $CuCl_4$ modes of vibration behave oppositely, showing hardening (by up to 6 $cm^{-1}$), but diminished band intensity. Accordingly, the $CuCl_4$ moieties get quantum confined in vdWs interaction dominates between the layers. Uniquely, as $t \rightarrow 150$ nm, also 177 $cm^{-1}$ band (Cl-Cu-Cl angle bending at a tetragonal $CuCl_4$ plane) adapts marginally enhanced intensity, likely on $CuCl_6^{4-}$ octagons are confined in $CuCl_4^{2-}$ (or $CuCl_4^{3-}$) tetragons. At t = 225 –150 nm (Figure 3e), the C-F stretching band (731 $cm^{-1}$) is vanished and the C-C ring breathing mode is stiffened from at 1002 $cm^{-1}$ (bandwidth $\Delta\nu_{1/2}$ = 10 $cm^{-1}$) in the starting $(3\text{-FBA})_2CuCl_4$ to at 1095 $cm^{-1}$ of a broader peak ($\Delta\nu_{1/2}$ = 105 $cm^{-1}$). Thus, the aromatic rings are quantum confined on the metal-ligand bonding is enriched at this critical stage. This specific vibration mode is not visible at intermediate states of the charge/spin orders, viz., damped on highly polar molecular chains (helicoids), as shown in the lattice images. Also, the N-H stretching mode loses intensity (Figure 3f) and ultimately disappears in loss-of energy in hydrogen bonding to the $CuCl_6^{4-}$ octagons, which are ultimately confined in square-planar $CuCl_4^{2-}$ and/or $CuCl_4^{3-}$ moieties,[42] showing duly stiffer phonons. A change of $Cu^{2+}$–coordination number at surface of (phenethylammonium)$_2CuCl_4$ is shown to exhibit photochromism.[40] These implications require studies of impacts of different OIHPs in harvesting light-induced pyro-photonic, thermo-chromic, and optoelectronic properties.

A survey scan of XPS (X-ray photoelectron spectroscopy) in Figure S4a confirms C, N, F, Cu and Cl are present in the (3-FBA)$_2CuCl_4$ composition. Two C1s bands are shown at binding-energy $E_b$ = 284.55 eV and 286.15 eV (Figure 4a) in two types C1s of C-C and C-N bonds, respectively. The N1s band of C-$NH_3$ heads at 401.55 eV (Figure 4b) contains a satellite band at 399.35 eV in a N-H---Cl (donor-acceptor) hydrogen bonding to the $CuCl_4$ units, which distorts the $CuCl_6^{4-}$ polygons.[35] The F atoms exhibit a single symmetric F1s band at 686.78 eV (Figure 4c), which signifies they do not hydrogen bond in (3-FBA)$_2CuCl_4$ is anchored in 2D layers. Seven Cu2p bands are shown in Figure 4d (marked at bars) over 930 $\rightarrow$ 965 eV in the $CuCl_4$ units are varied in valence states. The Cu–valence state can vary in Cl-polygons on Cu-sites in two major factors; (i) part of $Cu^{2+}$ is reduced to $Cu^+$ and (ii) that assume reduced $CuCl_6^{4-}$ –$CuCl_6^{5-}$ $\rightarrow$ $CuCl_4^{2-}$–$CuCl_4^{3-}$ co-ordinations in the forms of $Cu_2Cl_{12}^{9-}$ $\rightarrow$ $Cu_2Cl_{10}^{7-}$ , $Cu_2Cl_8^{5-}$ type oligomers. Multiple $Cu^{2+}$ co-ordinations were proposed earlier in similar OIHPs.[40] Thus, a major Cu2p$_{1/2,3/2}$ doublet $(CuCl_6)^{4-}$ arises in two bands at 954.08 eV (2p$_{1/2}$) and 934.28 eV (2p$_{3/2}$), with an intensity ratio $I_{1/2} = I_{3/2} = 0.51$ and a spin-orbit splitting $\Delta$ = 19.80 eV. A smaller $CuCl_4^{2-}$ type polytope shares a parallel doublet (a mirror image) displaced on its higher energy side by 1.45 eV ($\Delta$ = 19.60 eV). Two weaker bands (marked at stars) displaced at 963.51 eV (2p$_{1/2}$) and 942.50 eV (2p$_{3/2}$), with $\Delta$ = 21.01 eV eV), can be attributed to $Cu_2Cl_{12}^{9-}$/$Cu_2Cl_{10}^{7-}$ type oligomers. The last band at 945.10 eV ((2p$_{3/2}$) belongs further smaller $Cu_2Cl_9^{6-}$/$Cu_2Cl_8^{5-}$ type oligomers. Duly enhanced $E_b$ values of the Cu2p$_{1/2,3/2}$ XPS bands reveal the oligomers are quantum confined. Such oligomers are displaced in small dots in the HRTEM images (Figures 2d, e). A schematic diagram of the Cu2p$_{1/2, 3/2}$ energy levels (Figure 4e) highlights how that are pushed down (lower metallicity) below the Fermi-level ($E_F$) in quantum confined states of the oligomers. Pictograms in the right panel project the 2p$_{1/2, 3/2}$ bands in average harden in the $CuCl_{6-n}$ ($n = 0 \rightarrow 2$) polygons shrink in the smaller and smaller ($Cu^{2+} \rightarrow Cu^+$) oligomers. Mixed $CuCl_4$ polytopes cadged in 3-FBA type organic moiety lose inversion symmetry and function over multiple energy states of a high-entropy inorganic-organic hybrid heterostructure.

Heating and cooling a sample (3-FBA)$_2CuCl_4$, at 5 $Ks^{-1}$ in a 385 – 425 K range, in a differential scanning calorimeter (DSC) exhibit a reversible phase transition. An enthalpy $\Delta H_1$ = 1.135 J/g-K ($\Delta S_1$ = 2.852 mJ/g-K entropy) is stored in an endothermic peak at $T_1$ = 398 K in the heating (Figure 4f), and only its part of $\Delta H_2$ = (-) 0.235 J/g ($\Delta S_1$ = (-) 0.570 mJ/g-K) is released in an exothermic peak at $T_2 = 412$ K in the cooling. A large 79.3 % part of $\Delta H_1$ is absorbed (reorders of the small entities) in a thermal hysteresis, $\Delta T \equiv T_1 - T_2 \cong 14$ K. Both $\Delta H_1$ and $\Delta H_2$ are large values intrinsic of the first-order phase transition.[24,34] Also $\Delta T$-value is larger from (benzylammonium)$_2$PbCl$_4$,[30] with $T_1 = 438$ K and $T_2 = 433$ K, showing a due effect of the fluorination promotes the thermal hysteresis. The (3-FBA)$_2$CuCl$_4$ is thermally stable till 461 K in thermogravimetry (Figure S4b), viz., substantially stable well beyond the $T_C \rightarrow T_1 \sim$ 412 K shown in the electric dipoles release heat (in the DSC thermogram) in the disorder-order transition. Near room temperature, the dielectric permittivity $\varepsilon_r \sim 26$ is found to be well-stable (Figure 4g), with a little dielectric-loss, $\leq 0.02$, of a ferroelectric phase at $10^2$ to $2x10^6$ Hz frequencies. A weak signal is marked in a charge-order at nearly $5x10^5$ Hz frequencies, before the onset of $T_C$ point at frequency modulated $\varepsilon_r$ drops (up to 18) sharply and the dielectric–loss is raised sharply up to 0.8 in a paraelectric state. The $\varepsilon_r$–value exhibits a peak at $T_C \sim 413$ K (Figure S4c) as measured over temperatures at 50-100 kHz low frequencies. There is no apparent shift at these frequencies, as it arises in the ferro $\rightarrow$ paraelectric transition, and no much dielectric relaxations. This is superior $T_C$-value from lead-free (light-weight) molecular ferroelectrics.[34,47,48] A high $T_C$ with a large $\Delta S$-entropy (a result of chirality of a heterostructure of small entities) extends workability and stability of ferroelectrics for applications of electronic devices in extreme thermal conditions in the aerospace, automobiles, and energy industries.

## 2.2 Dynamics of (3-FBA)$_2$CuCl$_4$ domains of 2D-layers

The (3-FBA)$_2$CuCl$_4$ exhibits reversible thermochromic behaviour. As shown in Figure S5, its initial yellowish color at 298 K is turned dark brown at $T_C \sim 413$ K, and is redeemed yellowish at 298 K in successive heating and cooling cycles. The chirality of (3-FBA)$_2$CuCl$_4$ of chains/layers facilitates the ferroelectric domains to slide at the 2D-layers so that rapidly transfer energy metal $\leftrightarrow$ ligand at in situ charge order. As evidenced with the phonon dynamics, the $Cu^{2+}$ ions exhibit spin-flipping in the $^2D_{5/2}$ spin-orbit sates (Figure 3a) confined in the $CuCl_6^{4-}$ cadges, which readily swipe towards the layers (at a restoring-force induced at a thermal gradient, $(\partial T/\partial t) > 0$), as a torque (rotor) to rapidly transmit the energy. Thermo-/photo-chromic behaviour is shown in $Cu^{2+}$–based OIHPs, such as $(CH_3NH_3)_2CuCl_4$,[42] (PED)CuCl$_4$,[49] (BED)$_2$CuCl$_4$,[49] among many others, where PED = N-phenylethylene diamine, and BED = N-benzylethylene diamine). In situ induced pyro-/piezo-electricity feed the response in heat/light (or other stimuli) induced dynamics of inbuilt ferroelectric domains. The synergetic domains self-adjust to control a minimum free-energy of the system through the local symmetry, surface charge/spin order, vdWs interactions, dipole-dipole interactions, and phase order.[3,18] An inter-system $Cu^{2+} \rightarrow Cu^{+}$ charge/spin order on varied $Cu^{2+}/Cu^{+}$ co-ordinations in small $CuCl_{6-n}$ (n $\rightarrow$ 2) oligomers (shown in the XPS bands) would modulate the charge/spin dynamics. In this strategy, dynamics of polarized domains are studied with PFM images of $(3\text{-FBA})_2CuCl_4$ at room temperature by applying ac modulation bias to excite and order the dipoles at the itinerant states.[3,18] A self-adjustable material of duly large polarization thus can function as an effective piezo-/pyro-electric material. Accordingly, the polarization response is studied in analyzing electromechanical behavior of $(3\text{-FBA})_2CuCl_4$ sheets using the DART-PFM technique as follows.

The PFM images in Figure 5a present topology of $(3\text{-FBA})_2CuCl_4$, wherein switching spectroscopy is performed in PFM mode (SS-PFM) to get insight into the domain reversal process. The SS-PFM response in Figure 5b presents a piezoelectric butterfly loop and the ferroelectric hysteresis loop with 180° phase, which confer polarization reversibility of the ferroelectric domains. Out-of-plane PFM phase images in Figure 5c indicate the presence of domains (marked at arrows) oriented in two major distributions at ~ 30° (dark brown) and ~130° (yellowish). The corresponding bimodal distribution of dipoles are provided in Figure S6a. Similar domain patterns are also observed in the in-plane PFM phase images (Figure 5d) providing the response of the in-plane polarization process. However, the population of domains with 135° (yellowish) are significantly less (Figure S6b). The amplitude images taken in the out-of-plane and in-plane modes in Figures 5e, f provide involvement of strong nanoscale piezoelectric coupling. Thus, a piezoelectric coefficient of $d_{33}$ = 6.1 pm/V is calculated from the slope of the butterfly loop (Figure 5b), which is closer to a value 15 pm/V reported in high-grade ferroelectric poly(vinylidene) fluoride and copolymers[50,51.] A maximum $d_{33}$ = 42.0 pm/V is known in poled dipoles (up to 86%) in negative polarity-based electrospun nanofibers of poly(vinylidene-chlorotrifluoroethylene) copolymers. It is the best value known so far in the ferro-/piezo-electric polymers.[51] Thus, the PFM results confer the $(3\text{-FBA})_2CuCl_4$ layers form ferroelectric domains with reversible polarization.

The $(3\text{-FBA})_2CuCl_4$ having $Cu^{2+} \rightarrow Cu^{+}$ of varied co-ordinations strongly absorb light (Figure 6a) over 200 nm to 1500 nm, in which absorbance $\alpha_m$ is varied up to $0.34 \times 10^5$ $cm^{-1}$, and a deep absorption dip is emerged over 500 – 700 nm in reverse absorption process. A large absorption background prevails, $\alpha_m \leq 0.25 \times 10^5$ $cm^{-1}$, of small oligomers and chains scatter light and order in itinerant sates. Otherwise, $Cu^{2+}$ based OIHPs exhibit poor light absorption in visible-NIR regions.[40,52] $E_g$ = 2.10 eV is estimated from the Tauc plot of the UV absorption edge. C-$sp^2$ electrons (C-C rings), N-$2s^2$ nonbonding electrons (•$NH_3$ heads), and $3d^9$-$Cu^{2+}$ electrons (also $3d^{10}$-$Cu^+$ holes) at the $CuCl_4$ heads absorb UV-visible-NIR radiations. The chirality regulates them to exchange energy over correlated sates. As illustrated in the inset (Figure 6a), C-$sp^2$ electrons (hot) excited in $\pi_0 \rightarrow \pi^*$ states bypass energy to the subsequently $3d^9$-$Cu^{2+}$ states, which results in a deep 'valley'. It is a type of hole-burning.[53] This is a 'photochromic' effect of hot-spots radiate the excess energy. It can be used as an 'optical gate' on successive on-off irradiations of UV-visible light. A $10^4$ $cm^{-1}$ order of $\alpha_m$ is shown in halide perovskites,[17,54] useful for light absorbers, pyro-phototronic devices, medical therapy, and other implications. A closer view (Figure 6b) insights three $\pi_0 \rightarrow \pi_1^*$, $\pi_2^*$, $\pi_3^*$ bands (215, 245 and 295 nm peak positions) are mixed together in a wide contour of the C-C rings, superposed on two weaker bands at 390 nm and 465 nm of peak positions from the N-$2s^2$ nonbonding electrons.[55,56] The C-$sp^2$ electrons (in the ground state $\pi_0$) at C-H, C-F, and C-$NH_3$ bonds (at the C-C rings) differ in the excited states exhibiting three distinct $\pi_0 \rightarrow \pi_1^*$, $\pi_2^*$, $\pi_3^*$ bands. Likewise, two types $n_1/n_2 \rightarrow \pi_1^*$, $\pi_2^*$ bands arise in N-$2s^2$ nonbonding $n_1$ electrons on the C-$NH_3$ bonds, while those of $n_2$ belong to hydrogen bonded •$NH_3$ to the $CuCl_4$ heads, respectively. The $n_2 \rightarrow \pi_2^*$ transition loses energy on the hydrogen-bonding. Figure 6c projects transition-routes of C-$sp^2$ and $n_1/n_2$ electrons absorb light-energy in exciting to the excited states.

The electron density of states (DOS) of valence band (VB) are important to determine the optical properties are tailored in a hybrid organic–inorganic 3-FBA)$_2CuCl_4$ structure. In this context, we studied its UV-photoelectron spectrum (UPS), as given in Figure 6d, measured using a He–I emission line of energy hν = 21.22 eV) at a bias field of (-) 10 V, which shows the features of both primary and secondary photoelectrons. Consistent to the XPS, phonon, and electronic bands, two DOS maxima are shown at 16.31 eV (intense) and 9.72 eV (weak) from two type species of the inner and surface layers, respectively. The VB edge (zoomed in the inset) is displaced at $E_{vf}$ = (-) 1.75 eV from the $E_F$ level, and the secondary electron cut-off energy $E_{cf}$ = (-) 17.35 eV, which is used to determine value of the work-function $\varphi_w$ = 2.12 eV in a relation,[17]

$$\begin{aligned} w_{\emptyset} &= h\nu - E_{cf} \\ &\cong h\nu - E_{cf} - E_{vf} \end{aligned} \quad (1)$$

after subtracting the $E_{vf}$ value. Figure 6e represents a magnified UPS spectra near the valence band-edge for better visualization. An open P-E loop of polarization (P) is shown (Figure 6f) over applied electric fields, E $\rightarrow$ 20 kV/cm, intrinsic of ferroelectric phase at room temperature. A remnant polarization $P_r$ = 0.125 μC/cm$^2$, as observed at E = 18.75 kV/cm at 0.20 kHz frequency, is gradually diminished up to 0.043 μC/cm$^2$ on frequency is raised up to 0.5 kHz. The frequency supresses the loop at frequency induced metallicity. Primarily, the $CuCl_4$ inorganic phase stems ferroelectricity in a hybrid (3-FBA)$_2$CuCl$_4$ semiconductor. Although superior $P_r$ is reported at lead halides $PbX_4$ (Cl, Br, I) in OIHPs, [30,31,35] that are not favourable for light-weight electronics.

## 2.3 Pyro-phototronic (3-FBA)$_2$CuCl$_4$ device properties

The wide UV-visible-NIR light absorption, non-centrosymmetric structure (chiral), and ferroelectric properties of thin (3- FBA)$_2$CuCl$_4$ sheets are important to induce and regulate light-induced properties on the charge orders. To study the intricate coupling of the spontaneous polarization, optical, and semiconducting properties, we prepared a (3-FBA)$_2$CuCl$_4$ device with (Cr/Au) parallel electrode (channel length ~ 40 μm). Upon illuminating at $\lambda_{ex}$ ~ 455 nm 365 nm wavelengths by LEDs with a frequency of 0.1 Hz, a transient photocurrent $I_{sc}$ is induced, which is recorded under zero applied bias (Figure 6g, top panels). Four distinct stages of its response are shown in Figure 6h. The initial stage-1 before light-illumination describes the share of dark current. The net $I_{sc}$ is increased on lighting (stage-2) induces a photocurrent ($I_{Ph}$) of photogenerated charge carriers and a pyroelectric current ($I_{Py}$) is generated in situ on the temperature (T) is increased at a duly fast rate ($\frac{\partial T}{\partial t}$) > 0 (Figure 6g, bottom panels). A net $I_{sc} = I_{ph} + I_{py}$ value, as large as 60 nA (at 298 K $\rightarrow$ 301 K), is induced in UV-light illumination. This is 1-2 orders enhanced value than recently reported in organic and inorganic pyro-phototronic materials without applying external electric fields.[19,57] As the light-induced local thermal gradient saturates, ($\frac{\partial T}{\partial t}$) $\rightarrow$ 0, the $I_{Py}$ part gradually decreases and only the $I_{Ph}$ part prevails (stage-3). In the last and final stage-4, a reverse $I_{sc}$ develops in the light is turned-off and the device rapidly cools down, ($\frac{\partial T}{\partial t}$) < 0, leaving out only $I_{Py}$ of a reversed signal. With the symmetric Cr/Au electrodes and accounting for the absence of applied bias, the only built-in internal electric field persists, which allows discriminating the role of light-induced charge carriers in the spontaneous polarization from those of the ferroelectricity. Meanwhile, the electrodes are in close contact to the device and any change in polarization in temporal fluctuations leads to induce a due change in the bound charges at interfaces to the electrodes.

The light-illumination induce both charge-order and thermal gradient that lead to induce pyroelectric current expressed by,[5,10]

$$I_{Py} = P_c A \left(\frac{\partial T}{\partial t}\right) \tag{2}$$

where $P_c$ is the pyroelectric coefficient and A is the active area of the device. Thus, a distinct photocurrent response is regulated over the polarization changes in response to in situ light-induced ($\frac{\partial T}{\partial t}$), otherwise known as pyro-phototronic effect ($I_{Py+Ph}$). Since the $I_{Py}$ is related to the temporal temperature fluctuation, the resultant pyro-phototronic effect does depend on the change of instantaneous temperature as well. As portrayed in Figure 6g (bottom panel), the temperature change in the system during light illumination is monitored simultaneously. Increasing the LED power (used to illuminate the system) duly promotes the temporal change of the temperature ($\Delta T$), and in turn due increase in the net $I_{sc}$ output current. As depicted in Figure 6g, the net $I_{sc}$ demonstrates a direct proportionality to the light induced instantaneous $\Delta T$ values. Thus, a higher $\Delta T \rightarrow 3.2$ K is found at $\lambda_{ex} \sim 365$ nm (3.40 eV) illuminations than those at a higher $\lambda_{ex} \sim 455$ nm (2.73 eV) at almost same source power (~2 mW), which accounts for the effects of higher light-energy at shorter wavelengths. A tuneable photocurrent response at varied light-illuminations opens wide potentials to harvest waste thermal energy at photogenerated charge carriers. The source power of 1.4 mW, 12 mW, and 26 mW on illuminating at $\lambda_{ex} \sim 365$ nm, while 2 mW, 14 mW, and 27 mW, respectively, were used those at $\lambda_{ex} \sim 365$ nm for the pyro-phototronic measurements.

The light-induced pyro-phototronic response elucidates complex coupling in the thermo-/photo-chromocity, photon-phonon dynamics, and ferroelectricity of the thin $(3\text{-FBA})_2CuCl_4$ layers. The chirality seems to couple the coherent stimuli of vdWs, electrostatic, thermo-mechanical, and piezoelectric interactions to function together to exchange the energy from one form to others in the network channels. To get insight, dynamics of polar domains are studied with PFM phase images mapped under dark conditions (Figure 7a) and under illuminationn of blue light of $\lambda_{ex} \sim 455$ nm (Figure 7b). Distinct domains of yellowish contrasts (marked at arrows) are found to diminish under illumination (Figure 7b). As the illumination is removed, the polar domains revert to their initial positions (Figure 7c). A schematic diagram of the measurement set-up is described in Figure S6c. Similar features are shown (Figures 7d, e, f) in the PFM amplitude images. In situ induced thermal gradient, $(\frac{\partial T}{\partial t}) > 0$, on the illumination suppresses the polarization and induces $I_{Py}$ along with $I_{Ph}$, raising the net $I_{sc}$ in

stage-2 of the pyro-phototronic effect (Figure 6h). A diminished polarization is consistent with the suppression of ferroelectricity as the spontaneous polarization declines with in situ rising of temperature. The exposed area in the domains before and after removing the illumination is found to be almost the same of ~0.19 μm$^2$ and ~0.20 μm$^2$ respectively (while ~0.10 μm$^2$ during the illumination), which confer the domains are reversible in on-off illuminations. Earlier reports of photo-induced control of polar domains in organic and organic-inorganic hybrid halides is attributed to structural isomerization.[58-62] In absence of $(3\text{-FBA})_2CuCl_4$ photo-isomerization, here the thermal induced pyroelectricity causes the domain motion coupled to the light induced excitons.

Figures 7g- j present the effects of light-illumination on dynamics of domain visualized in PFM images (phase) of another $(3\text{-FBA})_2CuCl_4$ sample is exfoliated in multiple layers (t ~ 3 μm). The images are taken in dark and on illuminating the sample at successively decreased (increased light-energy) $\lambda_{ex}$ = 632 nm (red), 530 nm (green), 455 nm (blue), and 365 nm (UV), showing how the light-energy tunes the domains. A noteworthy is that a sufficient light-energy is absorbed at $\lambda_{ex}$ = 365 nm, 455 nm, and 530 nm, while there is no much absorption (consistent to the absorption bands in Figure 6a) at 632 nm (Figure 7h). Accordingly, the illumination supresses the polar domains in which the degree of suppression differs at different illuminations in order of $\lambda_{ex}$ values of 365 nm > 455 nm > 530 nm > 632 nm. The illumination at $\lambda_{ex}$ = 365 nm, viz. the highest light-energy 3.40 eV is used here, is highly effective to induce large pyroelectricity and in return duly large temporal $\left(\frac{\partial T}{\partial t}\right)$ gradients in the system. As a result, the polar domains are supressed most effectively at the UV-light illumination. Owing to less light-absorption, the illumination at $\lambda_{ex}$ = 632 nm (1.96 eV) causes effectively lower $\left(\frac{\partial T}{\partial t}\right)$ gradients, so to hardly suppress the polar domains. Figure 8a-c and 8d-f depicts the line profiles extracted from PFM phase and amplitude images in Figure 7a-c and Figure 7d-f, respectively (in dark and on after $\lambda_{ex}$ = 455 nm illumination). Figure 8a demonstrate quantified piezoresponse of domain motion at the polar surfaces. The domains in Figure 8b are shrunk over 8-11 μm (5-13 μm in the dark) on illuminating, and reverse domains are pinned in Figure 8c in opposite polarities over 5-13 μm of length scale along the surface. Similar piezoresponse is shown in the PFM amplitude in the domains shrinking. It could be observed that the PFM amplitude of the domains (yellowish) is increased to ~1.5 mV (from ~1.2 mV in dark) under illumination. This observation clearly indicates the photoinduced enhancement of electromechanical coupling in $(3\text{-FBA})_2CuCl_4$ is exfoliated in thin layers.[57] As shown at the bars in Figure 8g, the estimated area of the domains (yellowish) is found to be 0.193 $\mu m^2$, 0.093 $\mu m^2$, and 0.201 $\mu m^2$ in the respective PFM phase images. The distinct distributions of the domains present different phase polarizations on the light-induced charge orders. All these implications anticipate that the $(3\text{-FBA})_2CuCl_4$ that contains polar helices pave intricate coupling of light-induced excitons and ferro-/piezo-electricity in a pool of network channels. Here, no DC bias was used during the PFM imaging to avoid external influence of the intrinsic polarization.

## 3. CONCLUSIONS

A partially fluorinated halide perovskite $(3\text{-FBA})_2CuCl_4$ is developed using a partial F $\rightarrow$ H substitution at the aromatic rings, which paves excellent pyro-phototronic properties of a ferroelectric phase of a non-centrosymmetric orthorhombic crystal structure. It provides a model lead-free $Cu^{2+}$-based organic-inorganic hybrid 2D material (chiral structure) of polar molecular moieties and 2D-layers. District ferroelectric domains are shown at a nanoscale, with a reversible ferroelectric-to-paraelectric phase transition at $T_C$ = 398 K (412 K on cooling) in heating, a large 14 K thermal hysteresis, and a huge entropy-change $\Delta S_1 \equiv 2.852$ mJ/g-K out of a high entropy (chiral) structure. The chirality at molecular chains and 2D $(3\text{-FBA})_2CuCl_4$ layers leads to devise a high-entropy phase that favours the $T_C$ of such large $\Delta S_1$-value. The dynamics of polar domains is tuned at light-illuminations with LEDs in the UV-visible regions. The light-induced hot-spots induce pyroelectric current, but suppress the ferroelectric current, which results in a net value 60 nA of nearly two orders larger than known in organic and inorganic pyro-phototronic materials.[19,57] This is a strategy of coupled pyro-phototronics control and regulate the polarization (ferroelectricity) of the four-stage transient photo-response even if no any external bias field is applied. This is an efficient optical control of electric polarization is utilized to design next-generation self-powered optoelectronic devices, energy harvesters, and optical memories.

**Data Availability Statement**

The authors declare that the data supporting the findings of this study are available within the paper and in the Supporting Information.

**Supporting Information**

Supporting Information contains:

Experimental details, characterization techniques, crystallographic parameters, packing diagram of $(3\text{-FBA})_2CuCl_4$, bond length and bond angles of $(3\text{-FBA})_2CuCl_4$, XPS survey scan, TGA analysis, temperature dependent dielectric spectroscopy, thermochromism, population distribution of polar domains in PFM, schematic diagram of PFM measurement set-up under illumination.

**Accession Codes**

Deposition Numbers 2537903−2537904 contain the supplementary crystallographic data collected at low temperature and room temperature data, respectively. These data can be obtained free of charge via the joint Cambridge Crystallographic Data Centre (CCDC).

**Acknowledgements**

This work has been financially supported by a research grant of CRG-SERB, Govt. of India (CRG/2020/004306). ZM extends thanks to the University Grants Commission (UGC) for awarding the research fellowship (526979).

**Conflict of Interest**

The authors declare no conflicts of Interest.

**Data Availability Statement**

The data supporting the findings of the study and the corresponding structural data (CIF) file are provided in the Supporting information.

## REFERENCES

(1) Scott. J. F. Applications of modern ferroelectrics. *Science* **2007**, *315*, 954-959.
(2) You, Y. M.; Liao, W. Q. ; Zhao, D.; Ye, H. Y.; Zhang, Y.; Zhou, Q.; Niu, X.; Wang, J.; Li, P. F.; Fu, D. W., Wang, Z.; Gao, S.; Yang, K.; Liu, J. M.; Li, J.; Yan, Y.; Xiong, R. G. An organic-inorganic perovskite ferroelectric with large piezoelectric response. *Science* **2017**, *357*, 306-309.
(3) Song, X. -J.; Sun, W.; Zhou, L. -X.; Mao, W. –X.; Xu, H.-M.; Lan, J. -F.; Zhang, Y.; Zhang, H. -Y. Observation of ferroelectricity in carbapenem intermediates enables reactive oxygen species generation by ultrasound. *J. Am. Chem. Soc.* 2024, *146*, 32519-32528.
(4) Kumar, A.; Jain, A.; Naskar, S.; Ram, S.; Bera, C.; Mandal, D. Giant pyroelectric figure of merits in strain-engineered ferroelectric 2D-SnSe layered nanosheets: An efficient transient thermal energy harvester. *ACS Nano* **2025**, *19*, 19373-19383.
(5) Mishra, H. K.; Jain, A.; Saini, D.; Mondal, B.; Bera, C.; Ram, S.; Mandal, D. Ultrahigh pyroelectricity in monoelemental two-dimensional tellurium. *Phys. Rev. B* **2025**, *111*, 155436.

(6) Deng, W.; Zhou, Y.; Libanori, A.; Chen, G.; Yang, W.; Chen, J. Piezoelectric nanogenerators for personalized healthcare. *Chem. Soc. Rev.* **2022**, *51*, 3380-3435.
(7) Vijayakanth, T.; Shankar, S.; Finkelstein-Zuta, G.; Rencus-Lazar, S.; Gilead, S.; Gazit, E. Perspectives on recent advancements in energy harvesting, sensing and bio-medical applications of piezoelectric gels. *Chem. Soc. Rev.* **2023**, *52*, 6191.
(8) Wang, Q.; Tian, Y.; Yao, M.; Fu, J.; Wang, L.; Zhu, Y. Bimetallic organic frameworks of high piezovoltage for sono-piezo dynamic therapy. *Adv. Mater.* **2023**, *35*, e2301784.
(9) Zeng, X.; Liu, Y.; Weng, W.; Hua, L.; Tang, L.; Guo, W.; Chen, Y.; Yang, T.; Xu, H.; Luo, J.; Sun, Z. A molecular pyroelectric enabling broadband photo-pyroelectric effect towards self-driven wide spectral photodetection. *Nat. Commun.* **2023**, *14*, 5821.
(10) Saini, D.; Sengupta, D.; Mondal, B.; Mishra, H. K.; Ghosh, R.; Vishwakarma, P. N.; Ram, S.; Mandal, D. A spin-charge-regulated self-powered nanogenerator for simultaneous pyro-magneto-electric energy harvesting. *ACS Nano* **2024**, *18*, 11964-11977.
(11) Gao, X.; Zhang, X.; Yin, W.; Wang, H.; Hu, Y.; Zhang, Q.; Shi, Z.; Colvin, V. L.; Yu, W. W.; Zhang, Y. Ruddlesden–popper perovskites: synthesis and optical properties for optoelectronic applications. *Adv. Sci.* **2019**, *6*, 1900941.
(12) Liu, X.; Wang, S.; Long, P.; Li, L.; Peng, Y.; Xu, Z.; Han, S.; Sun, Z.; Hong, M.; Luo, J. Polarization-driven self-powered photodetection in a single-phase biaxial hybrid perovskite ferroelectric. *Angew. Chem.* **2019,** *131*, 14646-14650.
(13) Zhao, S.; Lan, C.; Li, H.; Zhang, C.; Ma, T. Aurivillius halide perovskite: A new family of two-dimensional materials for optoelectronic applications. *J. Phys. Chem. C* **2019**, *124*, 1788-1793.
(14) Li, M.; Xu, Y.; Han, S.; Xu, J.; Xie, Z.; Liu, Y.; Xu, Z.; Hong, M.; Luo, J.; Sun, Z. Giant and broadband multiphoton absorption nonlinearities of a 2D organometallic perovskite ferroelectric. *Adv. Mater.* **2020**, *32*, 2002972.
(15) Zhou, Y.; Chen, Z.; Wu, Z.; Shen, X.; Wang, J.; Zhang, J.; Sun, H. Hybrid improper ferroelectricity and magnetoelectric coupling in a two-dimensional perovskite oxide. *Phys. Rev. B* **2021**, *103*, 224409.
(16) Han, S.; Li, L.; Ji, C.; Liu, X.; Wang, G. E.; Xu, G.; Sun, Z.; Luo, J. Visible-photoactive perovskite ferroelectric-driven self-powered gas detection. *J. Am. Chem. Soc.* **2023**, *145*, 12853-12860.
(17) Mallick, Z.; Naskar, S.; Ram, S.; Mandal D. Light-regulated pyro-phototronic effects in a perovskite $Cs_2SnI_6$-reinforced ferroelectric polymer hybrid nanostructure. *Mater. Horiz.* **2025**, *12*, 1532-1546.
(18) Cui, X.; Ruan, Q.; Zhuo, X.; Xia, X.; Hu, J.; Fu, R.; Li, Y.; Wang, J.; Xu, H. Photothermal nanomaterials: a powerful light-to-heat converter. *Chem. Rev.* **2023**, *123*, 6891-6952.
(19) Dang, P.; Zhang, T.; Liu, Z.; Sun, S.; Wang, Y.; Zhang, F.; Wang, K.; Pan, X.; Lu, B.; Zhu, L.; Ye, Z.; Jiang, J. Tunable pyro-phototronic effect by polar interface engineering in $Ga_2O_3$. *Adv. Funct. Mater.* **2025**, *36,* 2504294.

(20) Wang, Z.; Yu, R.; Pan, C.; Li, Z.; Yang, J.; Yi, F.; Wang, Z. L. Light-induced pyroelectric effect as an effective approach for ultrafast ultraviolet nanosensing. *Nat. Commun.* **2015**, *6*, 8401.
(21) Yang, Q.; Guo, X.; Wang, W. H.; Zhang, Y.; Xu, S.; Lien, D. H.; Wang, Z. L. Enhancing sensitivity of a single ZnO micro-/nanowire photodetector by piezo-phototronic effect. *ACS Nano* **2010**, *4*, 6285-6291.
(22) Dai, Y. J.; Wang, X. F.; Peng, W. B.; Xu, C.; Wu, C. S.; Dong, K.; Liu, R. Y.; Wang, Z. L. Self-powered Si/CdS flexible photodetector with broadband response from 325 to 1550 nm based on pyro-phototronic effect: an approach for photosensing below bandgap energy. *Adv. Mater.* **2018**, *30*, 1705893.
(23) Liu, J.; Liang, Z.; Chen, J.; Wang, S.; Pan, C.; Qiao, S. A New Strategy of coupling pyroelectric and piezoelectric effects for photoresponse enhancement of a $Cu(In,Ga)Se_2$ heterojunction photodetector. *Adv. Funct. Mater.* **2022**, *32*, 2208658.
(24) Zhang, Q.; Solanki, A.; Parida, K.; Giovanni, D.; Li, M.; Jansen, T. L. C.; Pshenichnikov, M. S.; Sum, T. C. Tunable ferroelectricity in Ruddlesden–Popper halide perovskites. *ACS Appl. Mater. Interfaces* **2019**, *11*, 13523-13532.
(25) Haldar, R.; Sarkar, U.; Kumar, A.; Mandal, D.; Shanmugam M. Inducing polar phase in poly(vinylidene fluoride) with a molecular ferroelectric copper (II) complex for piezoelectric energy harvesting. *Adv. Funct. Mater.* **2024**, *34*, 2407611.
(26) Ai, Y.; Chen, X. G.; Shi, P. P.; Tang, Y. Y.; Li, P. F.; Liao, W. Q.; Xiong, R. G. Fluorine substitution induced high $T_C$ of enantiomeric perovskite ferroelectrics: (R)- and (S)-3-(fluoropyrrolidinium)$MnCl_3$. *J. Am. Chem. Soc.* **2019,** *141*, 4474-4479.
(27) Hou, Y.; Wu, C.; Yang, D.; Ye, T.; Honavar, V. G.; Van Duin, A. C.; Wang, K. Two-dimensional hybrid organic–inorganic perovskites as emergent ferroelectric materials. *J. Appl. Phys.* **2020**, *128,* 060906.
(28) Wang, J.; Li, J.; Tan, Q.; Li, L.; Zhang, J.; Zang, J.; Tan, P.; Zhang, J.; Li, D. Controllable synthesis of two-dimensional Ruddlesden–Popper-type perovskite heterostructures. *J. Phys. Chem. Lett.,* **2017**, *8*, 6211-6219.
(29) Ji, C.; Wang, S.; Li, L.; Sun, Z.; Hong, M.; Luo, J. The first 2D hybrid perovskite ferroelectric showing broadband white-light emission with high color rendering index. *Adv. Funct. Mater.* **2019**, *29*, 1805038.
(30) Liao, W. -Q.; Zhang, Y.; Hu, C.-L.; Mao, J. -G.; Ye, H.-Y.; Li, P.-F.; Huang, S. D.; Xiong, R.-G. A lead-halide perovskite molecular ferroelectric semiconductor. *Nat. Commun.* 2015, *6*, 7338.
(31) Ye, H. -Y.; Liao, W.-Q.; Hu, C. -L.; Zhang, Y.; You, Y.-M.; Mao, J. -G.; Li, P. -F.; Xiong, R. -G. Bandgap engineering of lead-halide perovskite-type ferroelectrics. *Adv. Mater.* **2016**, *28*, 2579-2586.
(32) Sha, T. -T.; Xiong, Y. -A.; Pan, Q.; Chen, X.-G.; Song, X. -J.; Yao, J.; Miao, S. -R.; Jing, Z. -Y.; Feng, Z. -J.; You, Y. -M.; Xiong, R. -G. Fluorinated 2D lead iodide perovskite ferroelectrics. *Adv. Mater.* **2019**, *31*, 1901843.

(33) Zhang, H. Y.; Zhang, Z. X.; Song, X. J.; Chen, X. G.; Xiong, R. G. Two-dimensional hybrid perovskite ferroelectric induced by perfluorinated substitution. *J. Am. Chem. Soc.* **2020,** *142*, 20208-20215.
(34) Tang, Y. Y.; Xie, Y.; Zeng, Y. L.; Liu, J. C.; He, W. H.; Huang, X. Q.; Xiong, R. G. Record enhancement of phase transition temperature realized by H/F substitution. *Adv. Mater.* **2020,** *32*, 2003530.
(35) Chen, X. G.; Song, X. J.; Zhang, Z. X.; Zhang, H. Y.; Pan, Q.; Yao, J.; You, Y. M.; Xiong, R. G. Confinement-driven ferroelectricity in a two-dimensional hybrid lead iodide perovskite *J. Am. Chem. Soc.* **2020,** *142*, 10212-10218.
(36) Zhang, Y.; Song, X. J.; Zhang, Z. X.; Fu, D. W.; Xiong, R. G. Piezoelectric energy harvesting based on multiaxial ferroelectrics by precise molecular design. *Matter* **2020,** *2*, 697-710.
(37) Guo, J.; Tao, L.; Xu, X.; Hou, L.; Nan, C. W.; Du, S.; Chen, C.; Ma, J. Rhombohedral R3 phase of Mn-doped $Hf_{0.5}Zr_{0.5}O_2$ epitaxial films with robust ferroelectricity. *Adv. Mater.* **2024**, *36*, 2406038.
(38) Guo, M.; Guo, C.; Han, J.; Chen, S.; He, S.; Tang, T.; Li, Q.; Strzalka, J.; Ma, J.; Yi, D.; Wang, K.; Xu, B.; Gao, P.; Huang, H.; Chen, L. -Q.; Zhang, S.; Lin, Y. -H.; Nan, C. -W.; Shen, Y. Toroidal polar topology in strained ferroelectric polymer. *Science* **2021**, *371*, 1050-1056.
(39) Caretta, A.; Miranti, R.; Havenith, R. W. A.; Rampi, E.; Donker, M. C.; Blake, G. R.; Montagnese, M.; Polyakov, A. O.; Broer, R.; Palstra, T. T. M.; van Loosdrecht, P. H. M. Low-frequency Raman study of the ferroelectric phase transition in a layered-based organic-inorganic hybrid. *Phys. Rev. B* **2014**, *89*, 024301.
(40) Groeneveld. B.; Duim, H.; Kahmann, S.; De Luca, O.; Tekelenburg, E. K.; Kamminga, M. E.; Protesescu, L.; Portale, G.; Blake, G. R.; Rudolf, P.; Loi, M. A. Photochromism in Ruddlesden–Popper copper-based perovskites: a light-induced change of coordination number at the surface. *J. Mater. Chem. C* **2020**, *8*,15377-15384.
(41) Dhanabalan, B.; Leng, Y-C.; Biffi, G.; Lin, M. -L.; Tan, P. -H.; Infante, I.; Manna, L.; Arciniegas, M. P.; Krahne, R. Directional anisotropy of the vibrational modes in 2D-layered perovskites. *ACS Nano* **2020**, *14*, 4689-4697.
(42) Elattar, A.; Suzuki, H.; Mishima, R.; Nakao, K.; Ota, H.; Nishikawa, T.; Inoue, H.; Kyaw, A. K. K.; Hayashi, Y. Single crystal of two-dimensional mixed-halide copper-based perovskites with reversible thermochromism. *J. Mater. Chem. C* **2021**, *9*, 3264-3270.
(43) Savi, L.; Celada, L.; Huu, D. K. A. P.; Chiesa, A. ; Carretta, S.; Painelli, A. Chirality-induced spin selectivity: A minimal model. *J. Phys. Chem. Lett.* **2025**, *16*, 9107-9115.
(44) Yadav, R. A.; Ram, S.; Shanker, R.; Singh, I. S. Vibrational spectrum, force field calculations, thermodynamic functions and barrier to internal rotation for benzoyl fluoride. *Spectrochim. Acta.* **1987**, *43A*, 901-909.
(45) Zhao, H.; Fu, H.; Hu, Z.; Fu, Q.; Tao, H.; Weng, J.; Xiong, L.; Cheng, Z. Magnetic hybrid organic–inorganic perovskite $(CH_3NH_3)_2XCl_4$ (X = Mn, Cu, Co) crystals. *Cryst. Eng. Comm.* **2021**, *23*, 5208-5213.

(46) Spirito, D.; Asensio, Y.; Hueso, L. E.; García, B. M. Raman spectroscopy in layered hybrid organic-inorganic metal halide perovskites. *J. Phys. Mater.* **2022**, *5*, 034004.
(47) Peng, Y.; Bie, J.; Liu, X.; Li, L.; Chen, S.; Fa, W.; Wang, S.; Sun, Z.; Luo, J. Acquiring High-$T_C$ layered metal halide ferroelectrics via cage-confined ethylamine Rotators. *Angew. Chem. Int. Ed.* **2021**, *60*, 2839-2843.
(48) Chen, Q.; Jiang, H.; Fan, Y.; Li, Z.; Ye, H.; Yao, Y.; Chen, S.; Ji, C.; Zhang, S.; Luo, J. High-$T_C$ Realization of lead-free halide hybrid ferroelectrics via steric confinement modulation. *Adv. Funct. Mater.* **2023**, *33*, 2213964.
(49) Sun, B.; Liu, X. F.; Li, X. Y.; Cao, Y.; Yan, Z.; Fu, L.; Tang, N.; Wang, Q.; Shao, X.; Yang, D.; Zhang, H. L. Reversible thermochromism and strong ferromagnetism in two-dimensional hybrid perovskites. *Angew. Chem. Int. Ed.* **2020**, *59*, 203-208.
(50) Gupta, V.; Babu, A.; Ghosh, S. K.; Mallick, Z.; Mishra, H. K.; Saini, D.; Mandal, D. Revisiting δ-PVDF based piezoelectric nanogenerator for self-powered pressure mapping sensor. *Appl. Phys. Lett*. **2021**, *119*, 252902.
(51) Kumar, A.; Gupta, V.; Malik, P.; Ram, S.; Mandal, D. Electrospun polarity-controlled molecular orientation for synergistic performance of an artifact-free piezoelectric anisotropic sensor. *Mater. Horiz.* **2024**, *11*, 4424-4437.
(52) Cortecchia, D.; Dewi, H. A.; Yin, J.; Bruno, A; Chen, S.; Baikie, T.; Boix, P. P.; Grätzel, M.; Mhaisalkar, S.; Soci, C.; Mathews, N. Lead-free $MA_2CuCl_xBr_{4-x}$ hybrid perovskites. *Inorg. Chem.* **2016**, *55*, 1044-1052.
(53) Phule, A. D.; Ram, S.; Shinde, S. K.; Choi, J. H.; Tyagi, A. K. Negative optical absorption and up-energy conversion in dendrites of nanostructured silver grafted with α/β-poly(vinylidene fluoride) in small hierarchical structures. *J. Phys. Chem. Solids* **2018**, *115*, 254-264.
(54) Dey, A.; Ye, J.; De, A.; Debroye, E.; Ha, S. K.; Bladt, E.; Kshirsagar, A. S.; Wang, Z.; Yin, J.; Wang, Y.; Quan, L. N.; Yan, F.; Gao, M.; Li, X.; Shamsi, J.; Debnath, T.; Cao, M.; Scheel, M. A.; Kumar, S.; Steele, J. A.; Gerhard, M.; Chouhan, L.; Xu, K.; Wu, X.; Li, Y.; Zhang, Y.; Dutta, A.; Han, C.; Vincon, I.; Rogach, A. L.; Nag, A.; Samanta, A.; Korgel, B. A.; Shih, C.; Gamelin, D. R.; Son, D. H.; Zeng, H.; Zhong, H.; Sun, H.; Demir, H. V.; Scheblykin, I. G.; Mora-Seró, I.; Stolarczyk, J. K.; Zhang, J. Z.; Feldmann, J.; Hofkens, J.; Luther, J. M.; Pérez-Prieto, J.; Li, L.; Manna, L.; Bodnarchuk, M. I.; Kovalenko, M. V.; Roeffaers, M. B. J.; Pradhan, N.; Mohammed, O. F.; Bakr, O. M.; Yang, P.; Müller-Buschbaum, P.; Kamat, P. V.; Bao, Q.; Zhang, Q.; Krahne, R.; Galian, R. E.; Stranks, S. D.; Bals, S.; Biju, V.; Tisdale, W. A.; Yan, Y.; Hoye, R. L. Z.; Polavarapu, L. State of the art and prospects for halide perovskite nanocrystals. *ACS Nano* **2021**, *15*, 10775-10981.
(55) Mishra, A.; Ram, S. Surface enhanced optical absorption and photoluminescence in nonbonding electrons in small poly(vinylpyrrolidone) molecules. *J. Chem. Phys.* **2007**, *126*, 084902.
(56) Rajeswari, P. V.; Sharma, S. K.; Ram, S.; Pradhan, D. Nanoporous N/O:$sp^2$-C films functionalized at nonbonding electrons of a biogenic husk (green chili) with deep UV-visible light absorption-emission for photocatalysis and other applications. *Surf. Interfaces* **2023**, *38*, 102824.

(57) Arora, K.; Saini, D.; Naskar, S.; Sahoo, S. C.; Mandal, D.; Neelakandan, P. P. Harnessing the piezo-phototronic effect in flexible chiral organic single crystals for light-responsive tactile sensing. *J. Am. Chem. Soc.* **2025**, *147*, 25498-25507.

(58) Du, Y.; Huang, C. R.; Xu, Z. K.; Hu, W.; Li, P. F.; Xiong, R. G.; Wang, Z. X. Optical control of polarization switching in a single-component organic ferroelectric crystal. *J. Am. Chem. Soc.* **2021**, *143*, 13816-13823.

(59) Wang, Z. X.; Huang, C. R.; Liu, J. C.; Zeng, Y. L.; Xiong, R. G. Salicylideneaniline is a photoswitchable ferroelectric crystal. *Chem. Eur. J.* **2021**, *27*, 14831-14835.

(60) Liao, W. Q.; Deng, B. B.; Wang, Z. X.; Cheng, T. T.; Hu, Y. T.; Cheng, S. P.; Xiong, R. G. Optically induced ferroelectric polarization switching in a molecular ferroelectric with reversible photoisomerization. *Adv. Sci.* **2021**, *8*, 2102614.

(61) Lv, H. P.; Xu, Z. K.; Yu, H.; Huang, C. R.; Wang, Z. X. A photochromic organic–inorganic hybrid Schiff base metal halide ferroelectric. *Chem. Mater.* **2022**, *34*, 1737-1745.

(62) Du, Y.; Huang, C. R.; Xu, Z. K.; Hu, W.; Li, P. F.; Xiong, R. G.; Wang. Z. X. Photochromic single-component organic Fulgide ferroelectric with photo-triggered polarization response. *J. Am, Chem. Soc. Au* **2023**, *3*, 1464-1471.

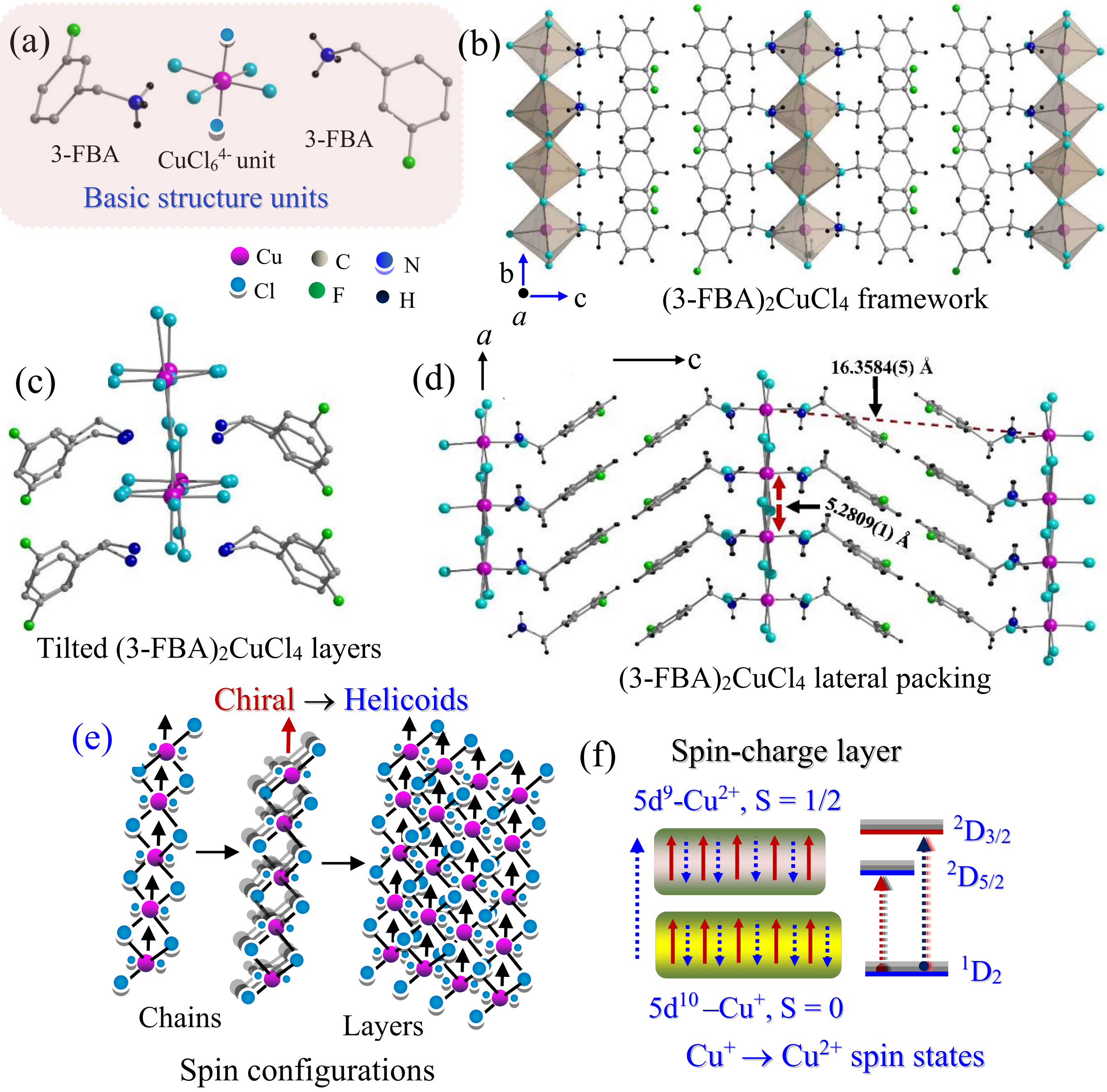

**Figure 1.** (a) Representation of the basic units of $(3\text{-FBA})_2CuCl_4$. Hydrogen atoms on the carbons are omitted for clarity. (b) Atomic packing of $(3\text{-FBA})_2CuCl_4$ along *a*-axis of $CuCl_4$ octahedra. (c) Symmetry breaking in $(3\text{-FBA})_2CuCl_4$ due to tilting of the layers with respect to one another and different positions of F-atoms in the 3FBAH moieties. (d) Atomic packing of $(3\text{-FBA})_2CuCl_4$ along b-axis showing the intra- and interlayer Cu…Cu distances. (e) $CuCl_6^{4-}$ chains/layers (chiral) with (f) a spin-charge layer.

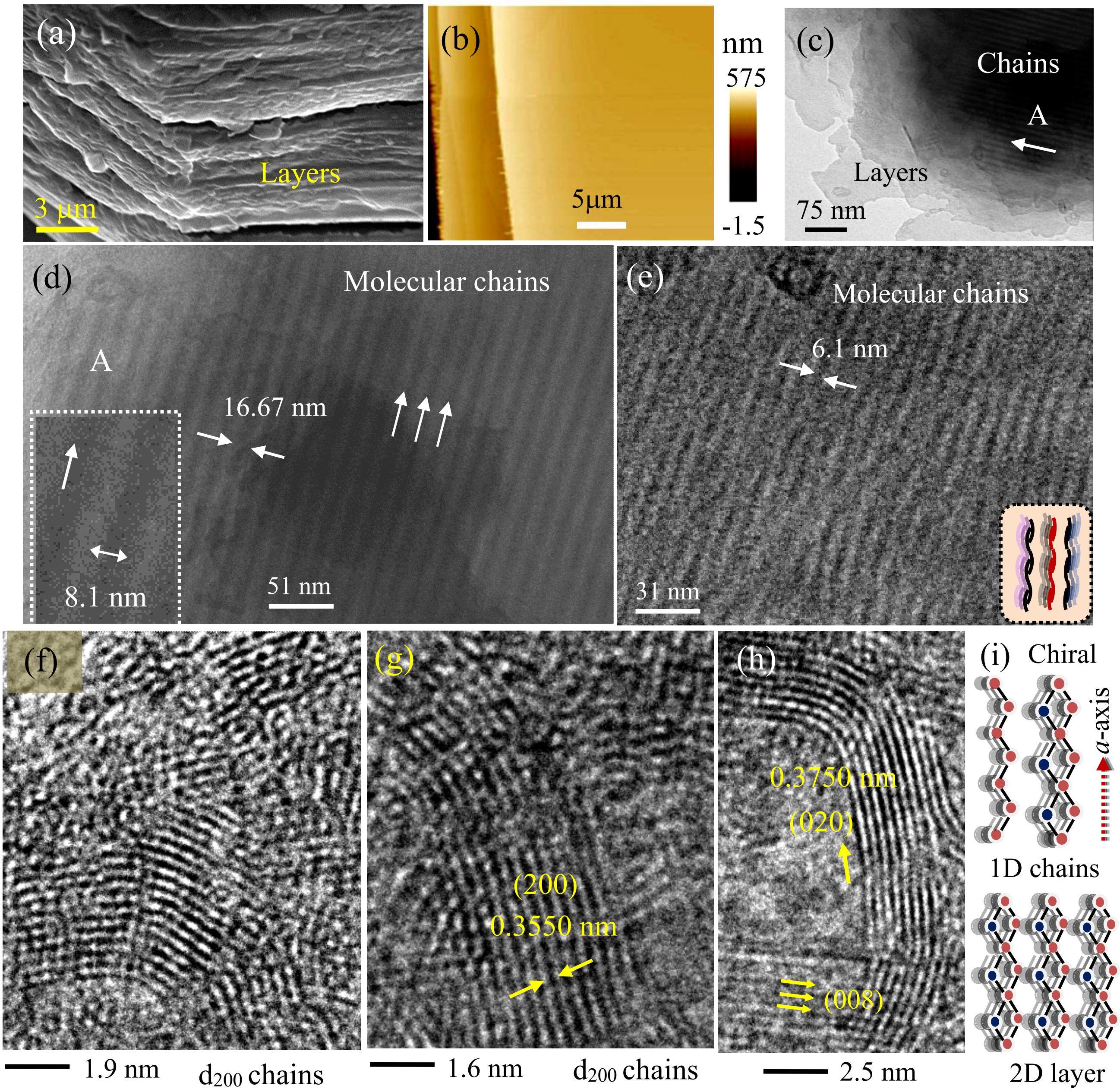

**Figure 2**. (a) FESEM, (b) AFM and (c) TEM images showing $(3\text{-FBA})_2CuCl_4$ is ordered in thin layers. (d) The molecular chains (a closer view in the corner) are wrapped in parallel strings (t = 8.3 nm), wherein (e) chirality (a model in the corner) is shown in refined strips (t = 6.1 nm). (f, g) Single molecular chains are ordered in a topology of toroid at $d_{200}$ = 0.3550 nm and (h) $d_{020}$ = 0.3750 nm (laid on $d_{008}$ = 0.4120 nm planes) atomic packing. (i) A model showing a single chiral chain can be wrapped in differed topologies.

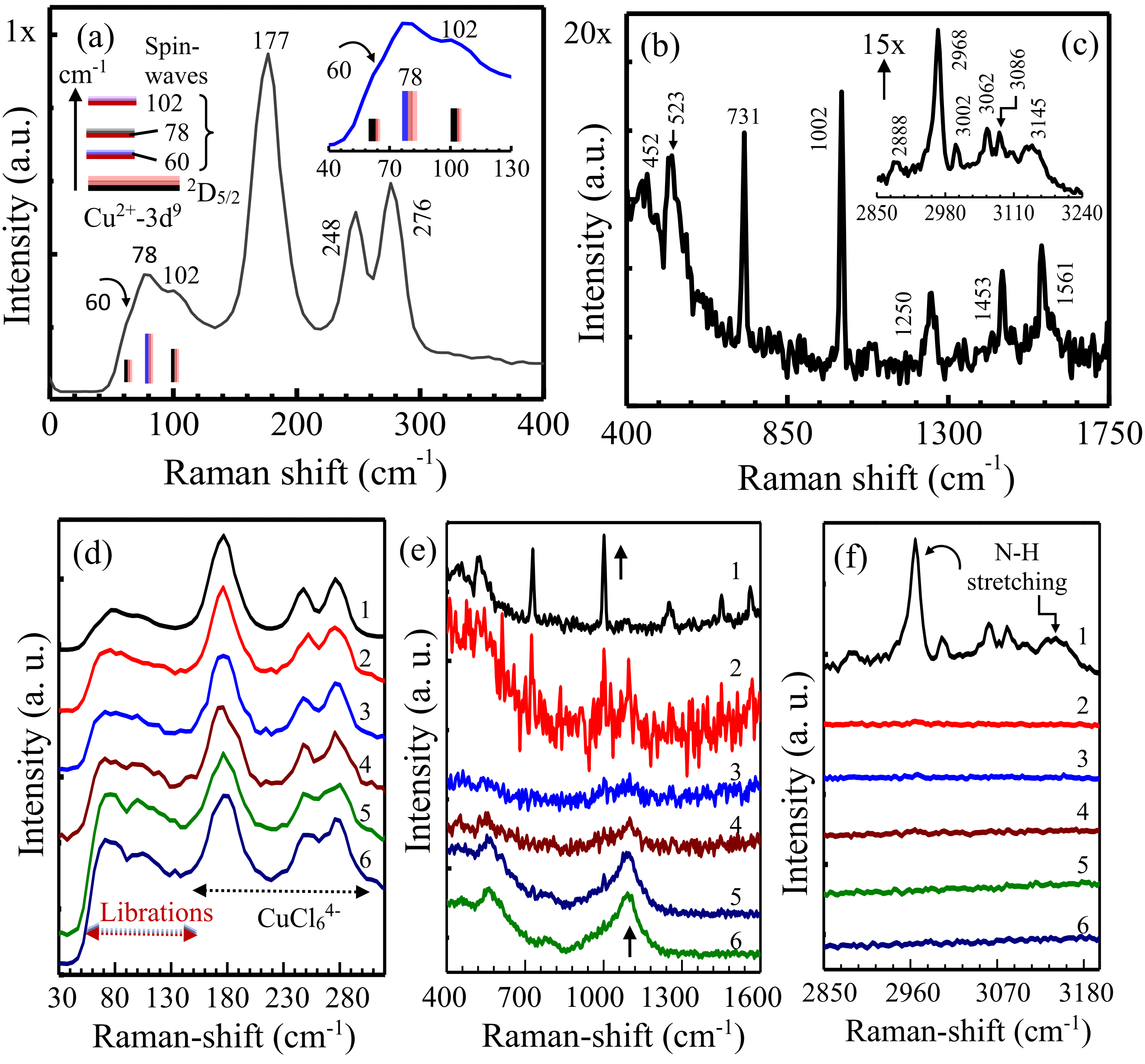

**Figure. 3.** Raman bands of $(3\text{-FBA})_2CuCl_4$; (a) 0 to 400 cm$^{-1}$, (b) 400 to 1750 cm$^{-1}$, and (c) 2850 to 3240 cm$^{-1}$, and the $(3\text{-FBA})_2CuCl_4$ is exfoliated in thin layers; (d) 20 to 300 cm$^{-1}$, (e) 400 to 1600 cm$^{-1}$, and (f) 2850 to 3200 cm$^{-1}$ regions. 1. Bulk and 2→ 6 thin layers of t = 600, 450, 350, 225, and 150 nm, respectively. The aromatic ring breathing mode is shifted 1002 → 1095 cm$^{-1}$ and N-H stretching mode is quenched in the thin layers.

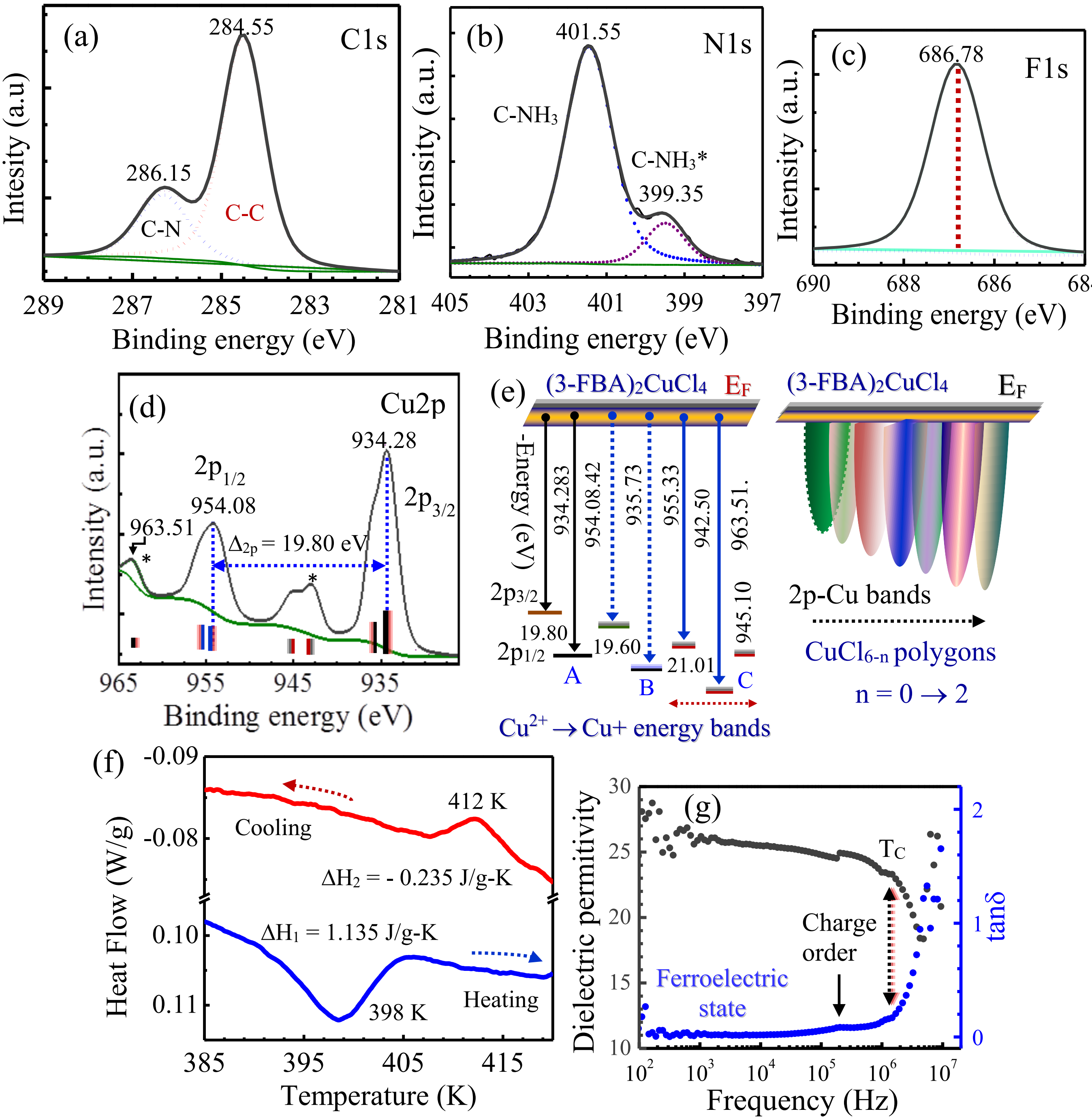

**Figure 4.** XPS (a) C1s, (b) N1s, (c) F1S and (d) $Cu2p_{1/2,3/2}$ bands in the (3-FBA)2CuCul4 sheets. (e) The $2p_{1/2,3/2}$ energy-levels derived from the XPS bands in (A) $CuCl_6^{4-}$, (B) $CuCl_4^{2-}$ and C) $Cu_2Cl_{12}^{9-} \rightarrow Cu_2Cl_{10}^{7-}$, $Cu_2Cl_8^{5-}$ type moieties. In the right panel, average $2p_{1/2}$ and $2p_{3/2}$ values are assumed to shift downwards (left to right) in smaller $CuCl_{6-n}$ polygons ($n = 0 \rightarrow 2$). DSC thermograms and (f) frequency dependent dielectric permittivity measured from the $(3\text{-FBA})_2CuCl_4$ sheets. *C-$NH_3$ hydrogen bonding to the $CuCl_4$ units.

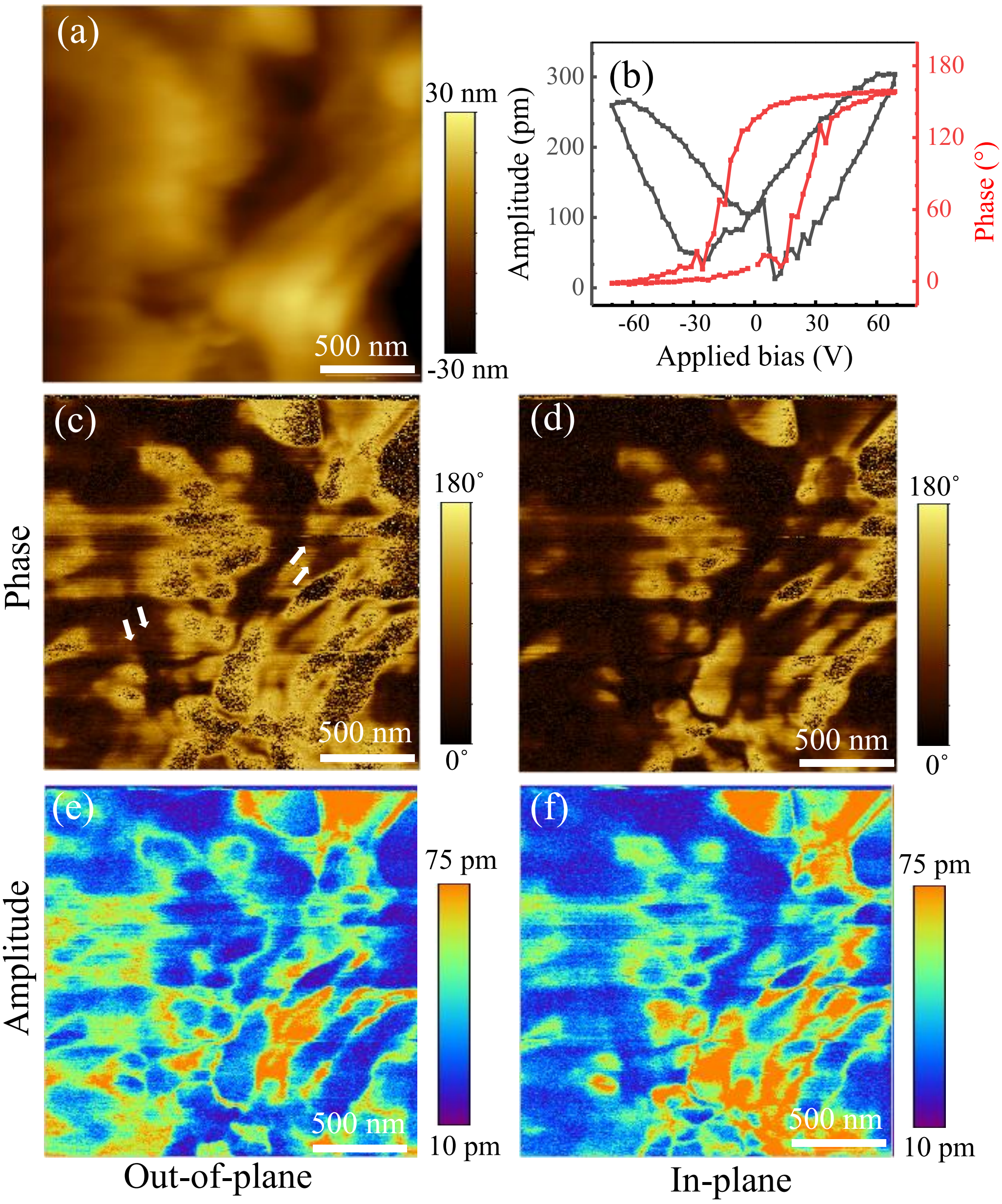

**Figure 5.** PFM investigation of the $(3\text{-}FBA)_2CuCl_4$ sheets. (a) Topography of the investigated area and (b) SS-PFM response indicating the presence of ferroelectric hysteresis loop with 180° phase reversal and piezoelectric butterfly loop. (c) The out-of-plane and (d) in-plane PFM phase responses. (e) Out-of-plane and (f) in-plane PFM amplitude responses.

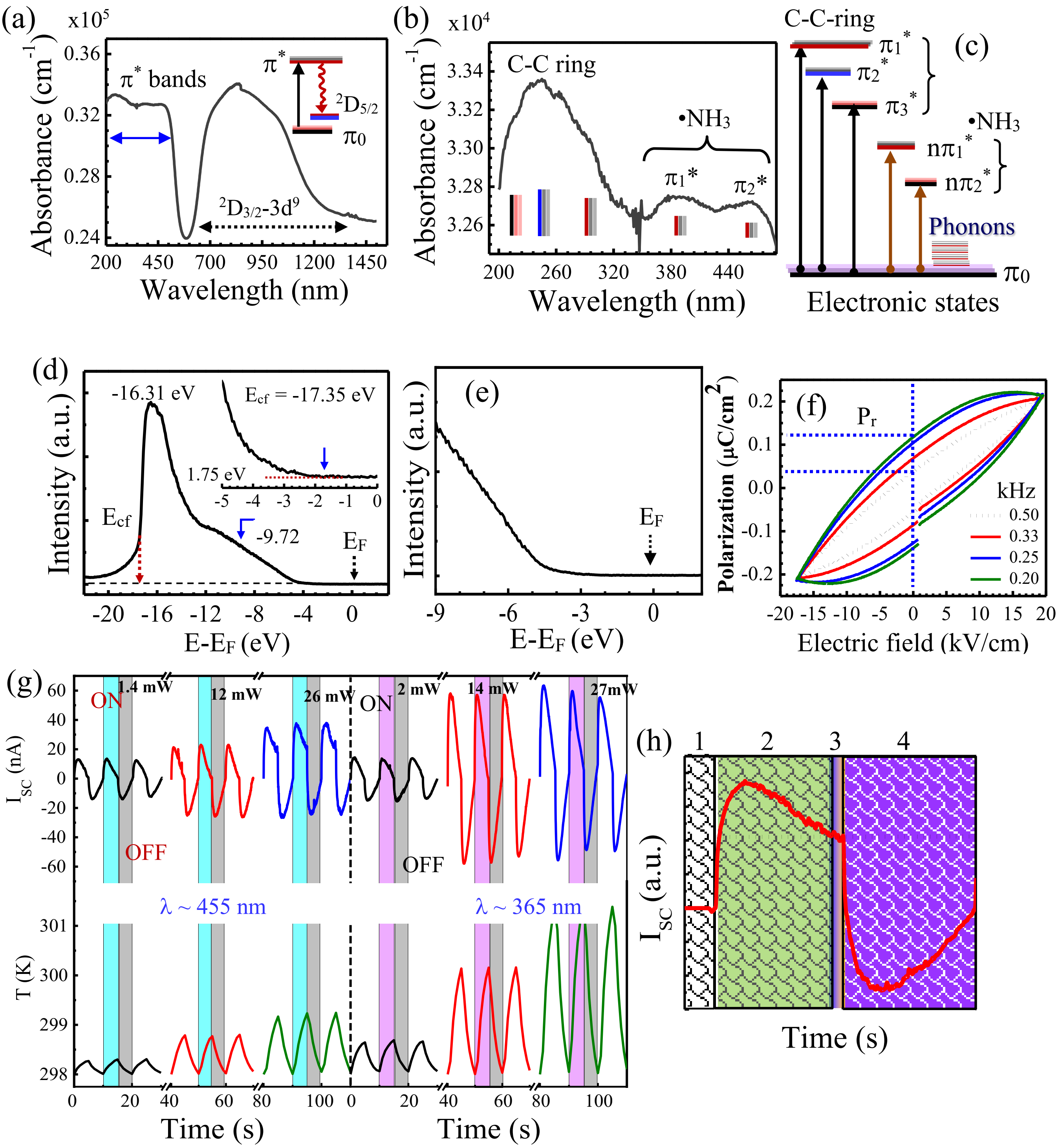

**Figure 6.** (a) Light absorption, (b) a closer view of C-C ring and C-$NH_3$ absorption bands, (c) schematic energy levels of the absorption bands, (d) UPS spectra of $(3\text{-FBA})_2CuCl_4$, with (e) parts magnified near the HOMO band-edge. and (f) P-E loops at selected frequencies of $(3\text{-FBA})_2CuCl_4$ sheets. (g) Photo-current response and change in temperature ($\Delta T$) on illuminating the sheets at $\lambda \sim 455$ nm and ~365 nm wavelengths of increased illumination power, showing linearly increased photocurrent with $\Delta T$, and (h) a schematic of four stages of the photocurrent resembling pyro-phototronic effect.

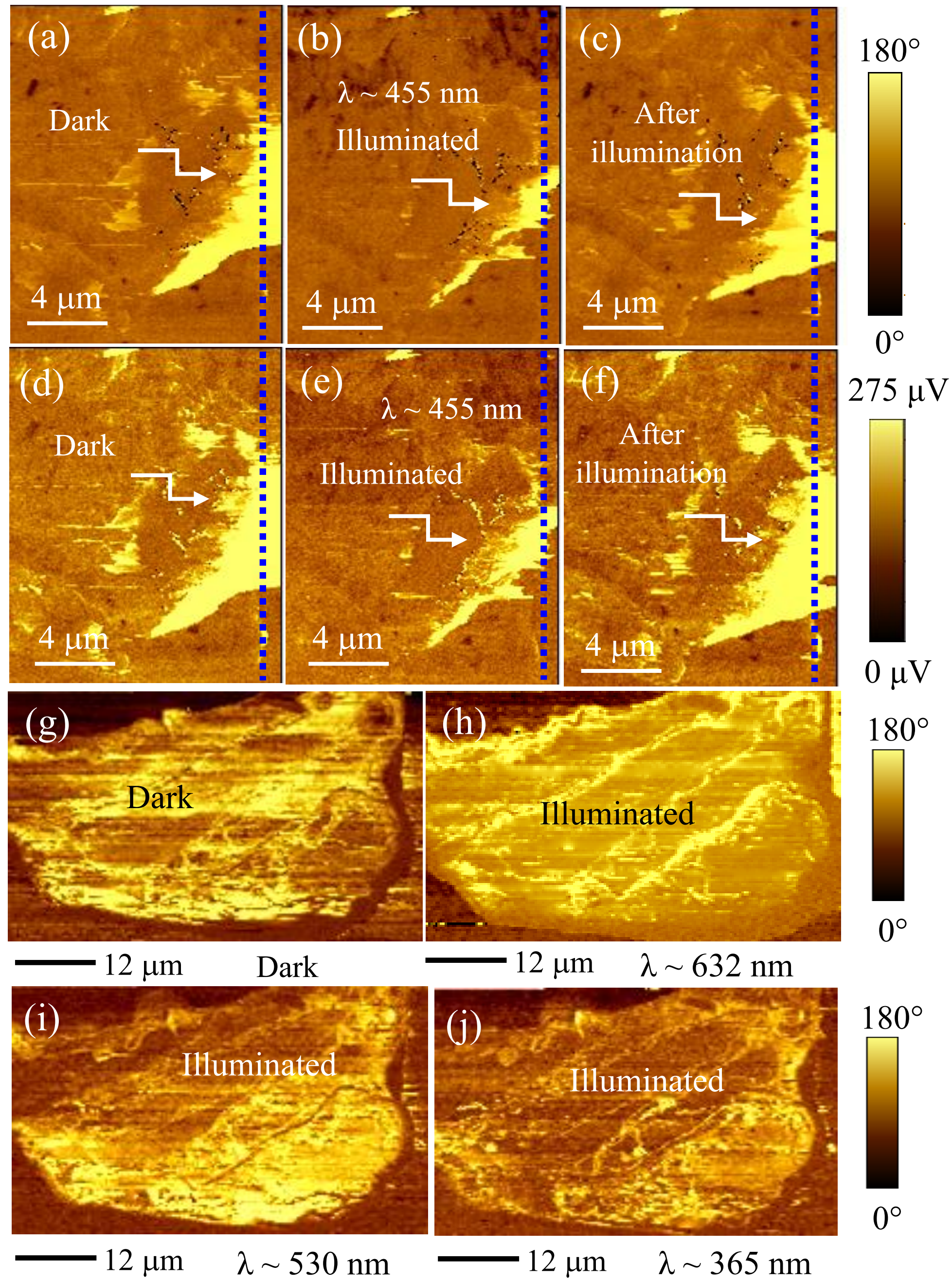

**Figure 7.** PFM images of the light induced polar domains from the thin $(3\text{-FBA})_2CuCl_4$ sheets in (a, b, c) phase and (d, e, f) amplitude modes in dark, illuminating at 455 nm, and after removing the illumination, respectively. The PFM images of another sheet in (g) dark and illuminating at (h) 632 nm, (i) 530 nm, and (j) 365 nm wavelengths.

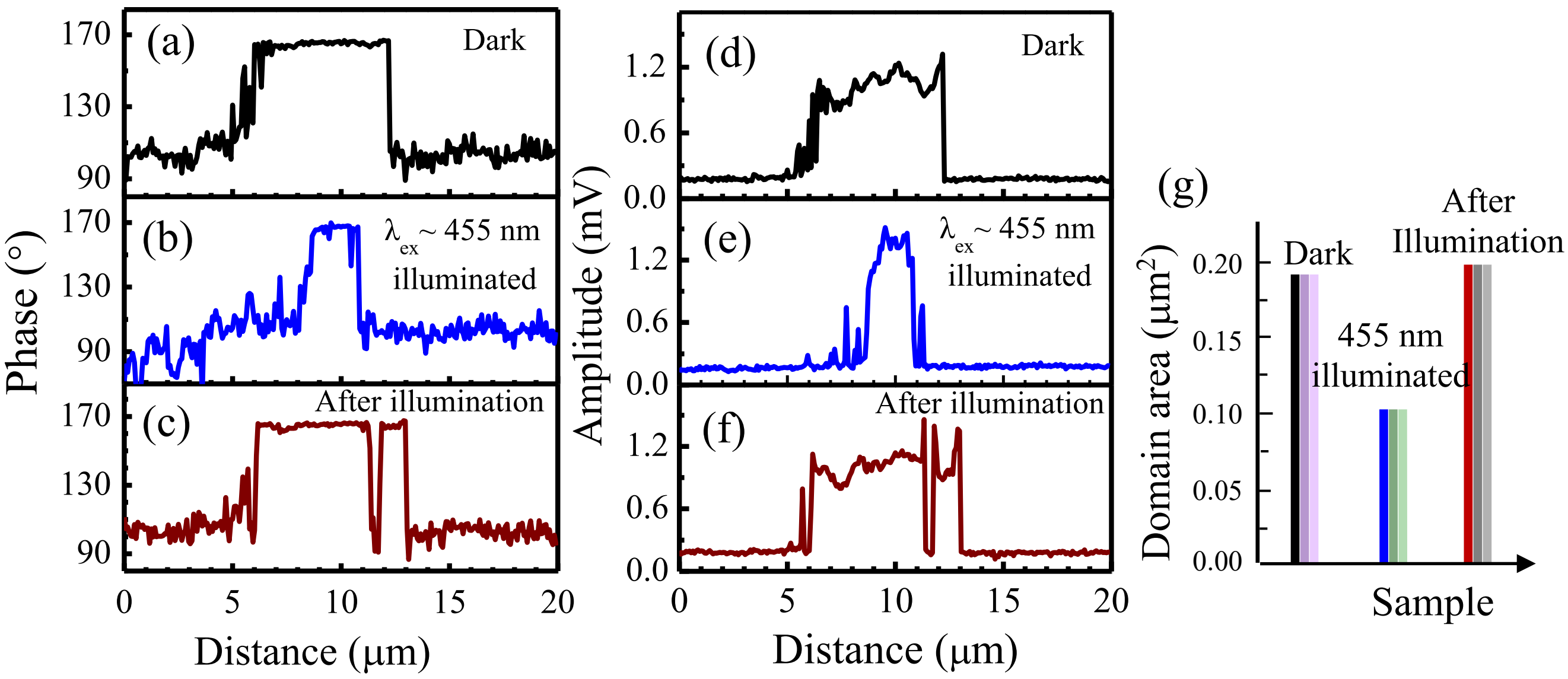

**Figure 8.** The line profiles of domain motions (projected along the dashed lines in Figure 7) from the PFM images of $(3\text{-FBA})_2CuCl_4$ sheets on different illumination conditions; (a, b, c) phase and (d, e, f) amplitude modes. (g) Projected areas of the yellow domains are marked during dark, under $\lambda_{ex} \sim 455$ nm, and after illumination (calculated from Figure 7a-c).

## Supporting Information

# Above Room-Temperature Phase Transition in Helicoidal 2D Halide Perovskite Enables Pyro-Phototronic Control

### Experimental Section

**Materials:** Copper (II) Chloride dihydrate (Sigma Aldrich); 3-Fluorobenzylamine (3-FBA) (TCI Chemicals); Ethanol (99.99%); Hydrochloric acid (HCl, 36 % w/w, aqueous solution). All the precursors were used without further purification.

**Crystal Growth:** Single crystals of 3-fluorobenzylammonium copper chloride $(3\text{-FBA})_2CuCl_4$, was crystallized following a slow evaporation technique. $CuCl_2.2H_2O$ (20 mM) was first dissolved in ethanol (2 mL) with slow stirring. Once completely dissolved, 3-FBA (40 mM) was added dropwise while stirring the solution. The produced precipitates were then dissolved by adding excess HCl to get a clear solution. Upon slow evaporation of the solvent in ambient temperature and pressure, yellow-coloured plate like crystals were obtained. The as grown crystals were used for further characterizations.

**Characterizations:** The single-crystal X-ray diffraction (SCXRD) data of the perovskite was collected on a Bruker diffractometer with Mo Kα radiation (λ = 0.71073 Å) at 150 K. Data reduction and the unit cell parameters were determined by using CrysAlisPro 1.171.38.43. With the help of Olex2 software and the SHELXL program, crystal data was solved by direct method and refined by the least square procedure. Powder XRD data of the perovskite was obtained using All non-hydrogen atoms were refined anisotropically and the positions of all hydrogen atoms were generated geometrically. Thermogravimetric analyses (TGA) were carried out on a Rigaku TGMS-ThemoMass Photo instrument with a heating rate of 5 K/min under nitrogen atmosphere. Differential Scanning Calorimetry (DSC) was performed using a Rigaku DSC Vesta instrument with heating and cooling rate of 5 K/min in aluminium crucibles under nitrogen atmosphere. The temperature dependence of dielectric permittivity of as grown crystals was measured with Keysight Impedance Analyzer E4990A in a parallel plate capacitor geometry. The macroscopic ferroelectric property was measured by Radiant Technologies Precision LC-II ferroelectric loop tracer. The nanoscale piezoelectric and ferroelectric properties were measured using piezoresponse force microscopy (PFM) in MFP-3D-Bio AFM in DART (Dual AC

Resonance Tracking) mode with Pt/Ir coated conductive AFM probe (SCM-PIT V2; k~ 3 N/m). by applying an AC modulating voltage of 2.5 V. The polarization domain dynamics under illumination (LED with λ 365 nm, 455 nm, 530 nm and 632 nm) was studied using Nanosurf FlexAFM (AFM probe: Multi75E-G; k~ 3 N/m). The chemical state of the perovskite was analysed using X-ray photoelectron spectroscopy (XPS) (K-alpha, Thermo Fisher Scientific, USA) equipped with monochromatic source of Al $K_{\alpha}$ (1486.7 eV). In the same instrument, the density of states in valence band and work-function of the system was determined using Ultra-violet photoelectron spectroscopy (UPS) with He-I UV source (21.22 eV). The optical absorbance was investigated in Agilent spectrophotometer. The pyroelectric and pyro-phototronic responses were recorded using Keithley 2450 source meter.

**Table S1.** Crystallographic parameters for $(3\text{-FBA})_2CuCl_4$

| **Parameter** | at 150K | at 300K |
|---|---|---|
| Empirical formula | $C_{14}H_{18}Cl_4CuF_2N_2$ | $C_{14}H_{18}Cl_4CuF_2N_2$ |
| Formula weight | 457.64 | 457.64 |
| Crystal system | Orthorhombic | Orthorhombic |
| Space group | $Pca2_1$ | $Pca2_1$ |
| a/Å | 7.2718(4) | 7.29551(4) |
| b/Å | 7.5384(5) | 7.5749(5) |
| c/Å | 32.329(2) | 32.5505(2) |
| α, β , γ /° | 90 | 90 |
| Volume/$Å^3$ | 1772.2(2) | 1798.82(2) |
| Z | 4 | 4 |
| $\rho_{calc} g/cm^3$ | 1.725 | 1.690 |
| $\mu/mm^{-1}$ | 1.852 | 1.825 |
| Temperature/K | 150.00(10) | 300.00(10) |
| $2\Theta_{max}$ | 50 | 50 |
| Radiation | $MoK_{\alpha}$ | $MoK_{\alpha}$ |
| λ [Å] | 0.71073 | 0.71073 |
| Reflns | 16890 | 15503 |
| Ind. reflns | 3904 | 3146 |
| Goodness-of-fit on $F^2$ | 1.094 | 1.087 |
| $R_1$ | 0.0712 | 0.0433 |
| $wR_2$ | 0.1730 | 0.1099 |

(a)

(b) (c)

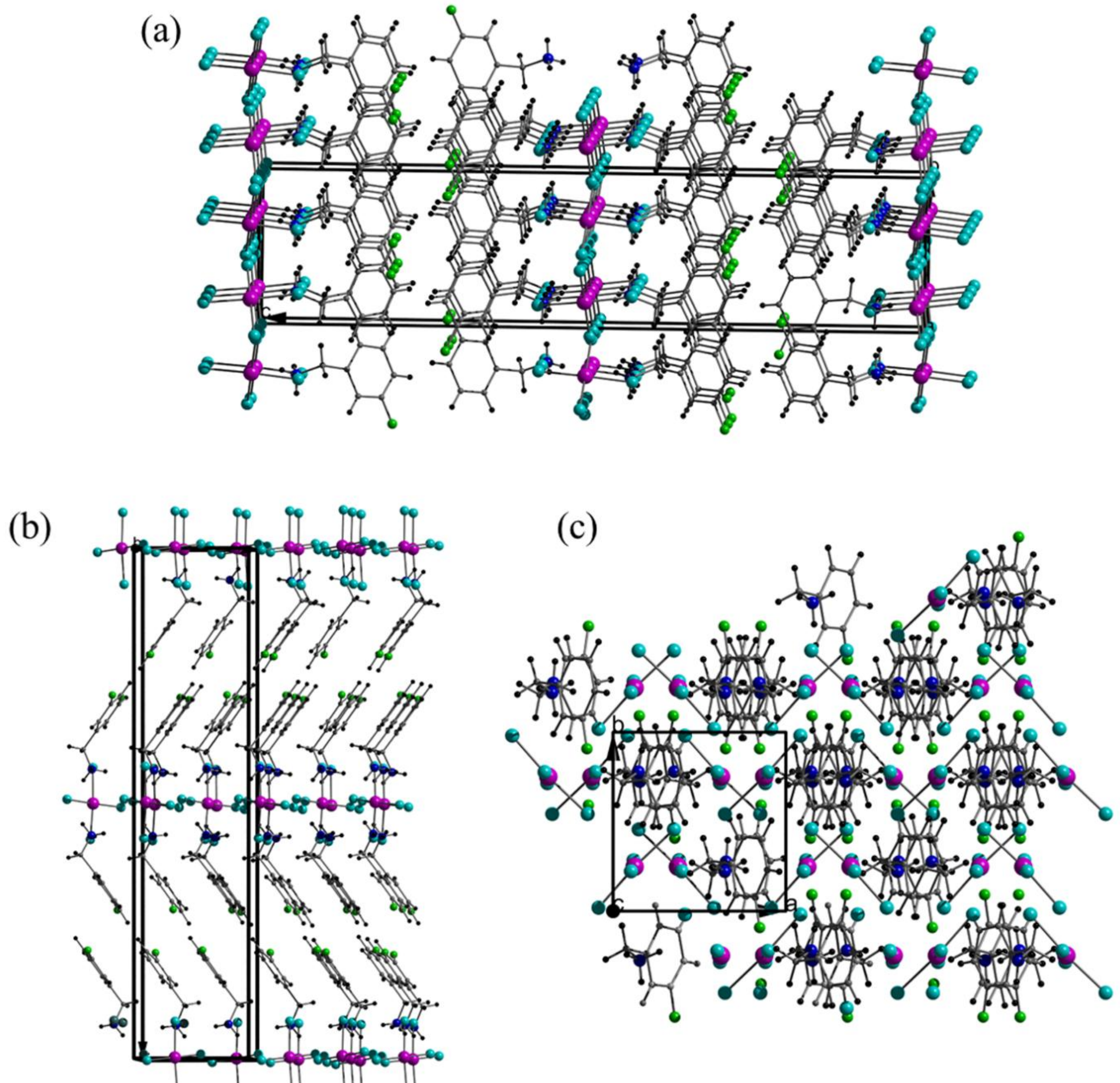

**Figure S1**. (a) Packing diagram of $(3\text{-FBA})_2CuCl_4$ along a-axis. Colour Code: Copper: magenta; Carbon: dark grey; Nitrogen: blue; Chlorine: turquoise, Fluorine: green; Hydrogen: black, (b-c) Packing diagram of $(3\text{-FBA})_2CuCl_4$ along b and c-axis.

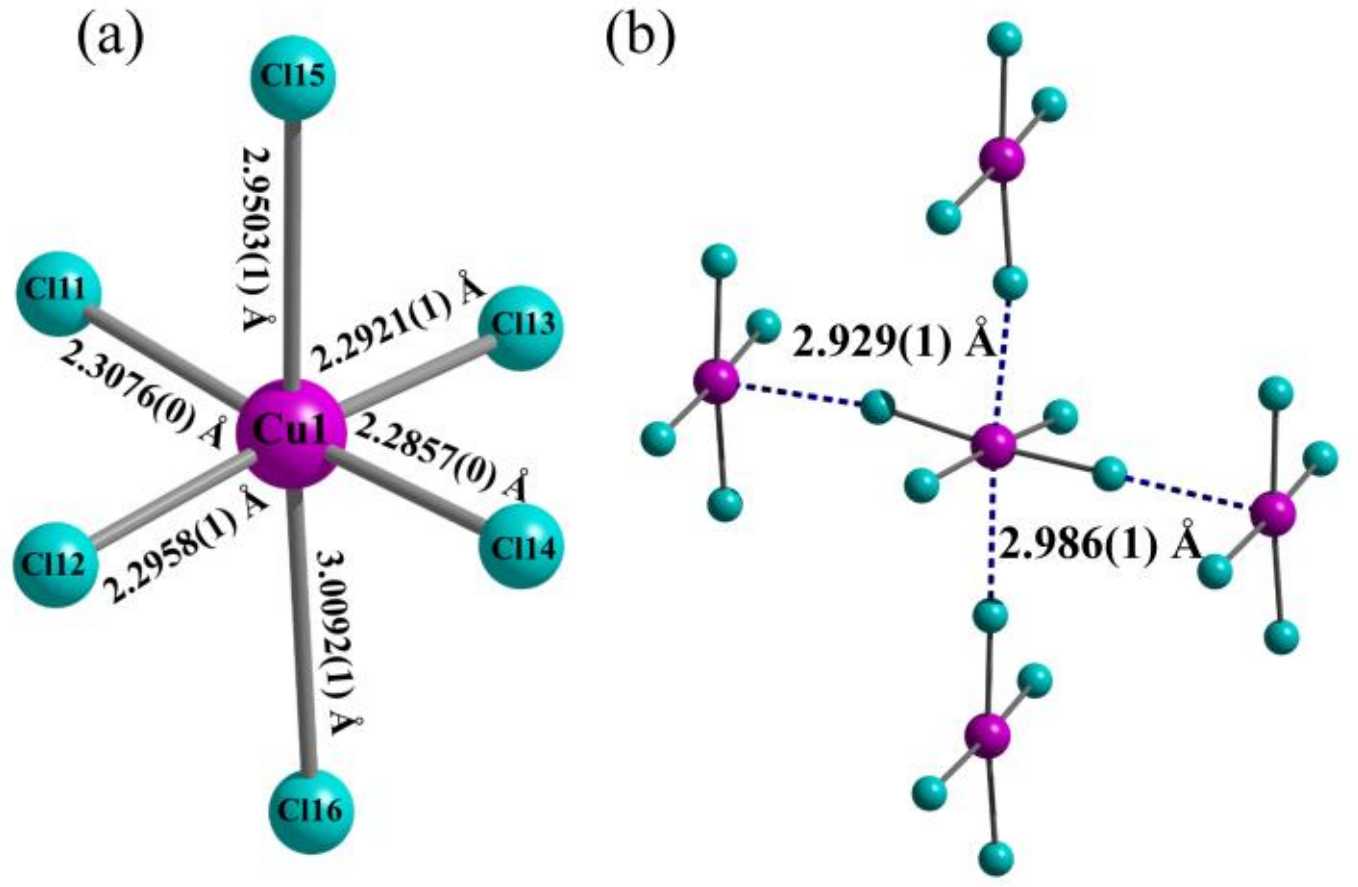

**Figure S2.** Selected bond lengths of $(3\text{-FBA})_2CuCl_4$**.**

**Table S2:** Selected bond angles of $(3\text{-FBA})_2CuCl_4$**.**

| Bond Angle | Value (°) |
|---|---|
| ∠Cl11-Cu1-Cl12 | 174.62(14) |
| ∠Cl11-Cu1-Cl13 | 88.59(18) |
| ∠Cl11-Cu1-Cl14 | 91.39(17) |
| ∠Cl12-Cu1-Cl13 | 89.61(18) |
| ∠Cl12-Cu1-Cl14 | 90.85(17) |
| ∠Cl13-Cu1-Cl14 | 174.89(13) |

**Table S3:** Atoms involved in intermolecular hydrogen bonding and its corresponding bond distances and bond angles in **1**.

| H-Bond Donor(D)- Acceptor(A) | D….A (Å) | ∠DHA (°) |
|---|---|---|
| N11-H11A…Cl11 | 3.331(15) | 167.7 |
| N11-H11A…Cl14 | 3.385(16) | 112.9 |
| N11-H11B…Cl11_$1 | 3.384(18) | 143.6 |
| N11-H11B…Cl13_$1 | 3.290(16) | 135.8 |
| N11-H11C…Cl14_$2 | 3.465(16) | 129.3 |
| N11-H11C…Cl13_$3 | 3.376(15) | 145.1 |
| N21-H21A…Cl14_$4 | 3.353(16) | 149.3 |
| N21-H21A…Cl13_$5 | 3.382(16) | 122.8 |
| N21-H21B…Cl14_$2 | 3.413(14) | 129.7 |
| N21-H21B…Cl12 | 3.327(15) | 140.3 |
| N21-H21C…Cl13 | 3.394(16) | 104.7 |
| N21-H21C…Cl12_$5 | 3.360(14) | 160.7 |
| C15-H15…F21_$6 | 3.44(3) | 138.6 |
| C15-H15…F11_$1 | 3.39(3) | 110.5 |
| C25-H25…F21_$7 | 3.47(2) | 105.8 |
| C25-H25…F11_$8 | 3.44(2) | 142.8 |

$1 = x+0.5, -y+1, z; $2= x-0.5, -y+1, z; $3 = x, y-, z; $4 = x-1, y, z; $5 = x-0.5, -y+2, z; $6 = -x+1.5, y-1, z-0.5; $7= x+0.5, -y+2, z; $8= -x+1.5, +y, z+0.5

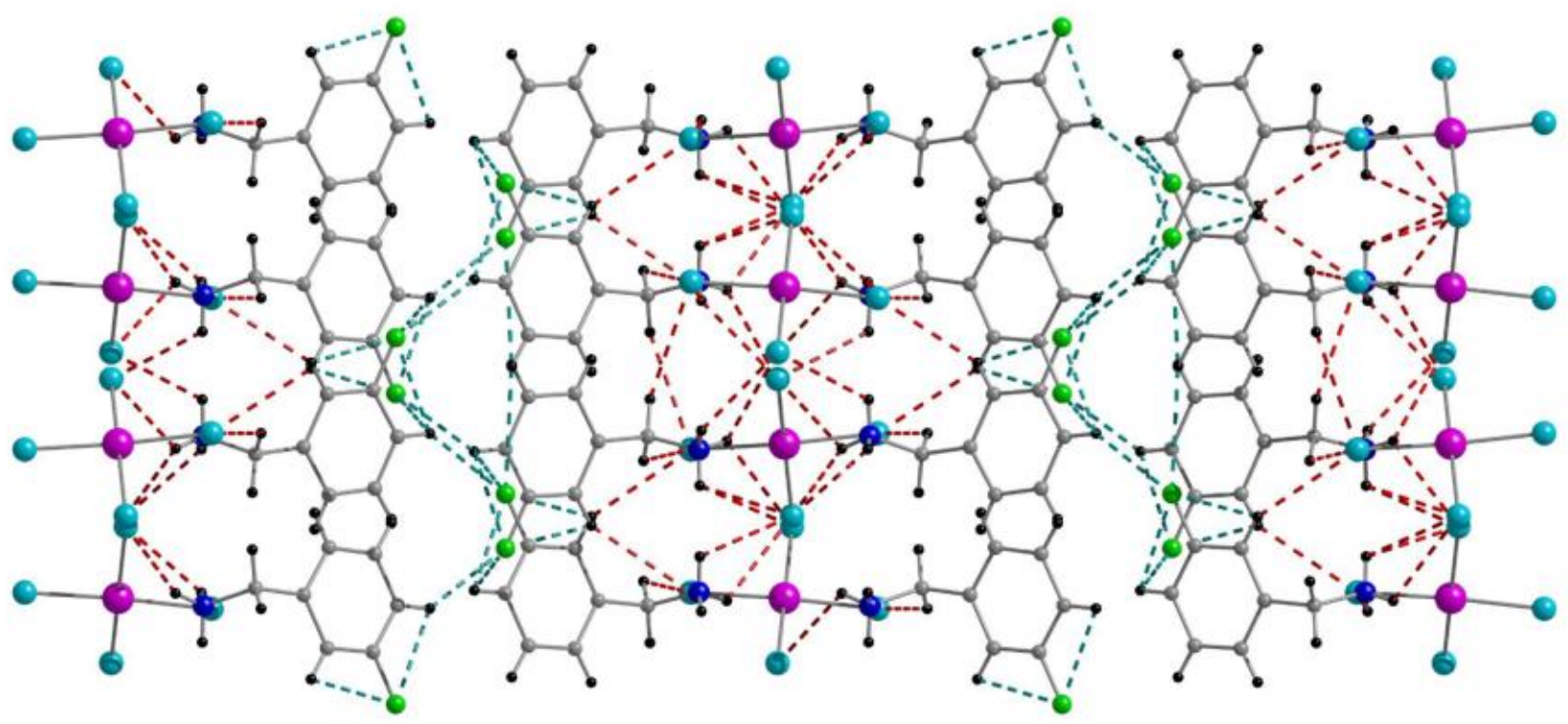

**Figure S3**. Inter and intramolecular classical N-H…Cl hydrogen bonding between H atoms of amine and the Cl atoms of the $CuCl_4$ moiety represented by dark red dotted lines and C-H…F hydrogen bonding represented by blue dotted lines.

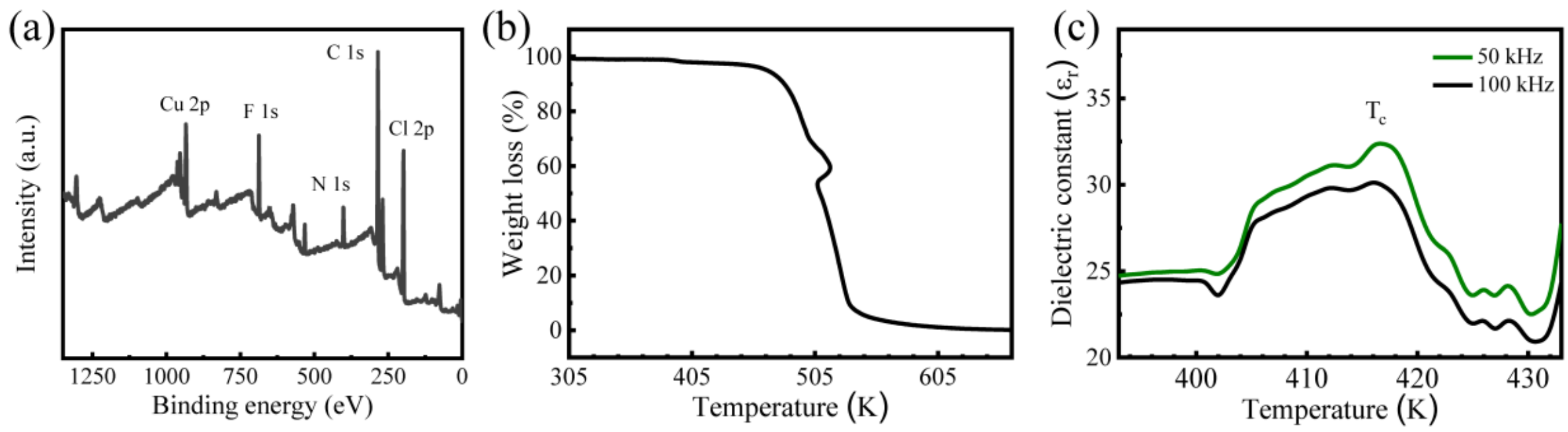

**Figure S4.** (a) XPS survey scan of $(3\text{-FBA})_2CuCl_4$ confirming the presence of elements as C, N, F, Cu and Cl. (b) The TGA analysis indicates the stability of the perovskite till the temperature of 461K, beyond which it decomposes. (c) Temperature dependent dielectric spectroscopy at 50 and 100 kHz indicating a ferroelectric to paraelectric phase transition around ~413 K, consistent with the DSC thermogram.

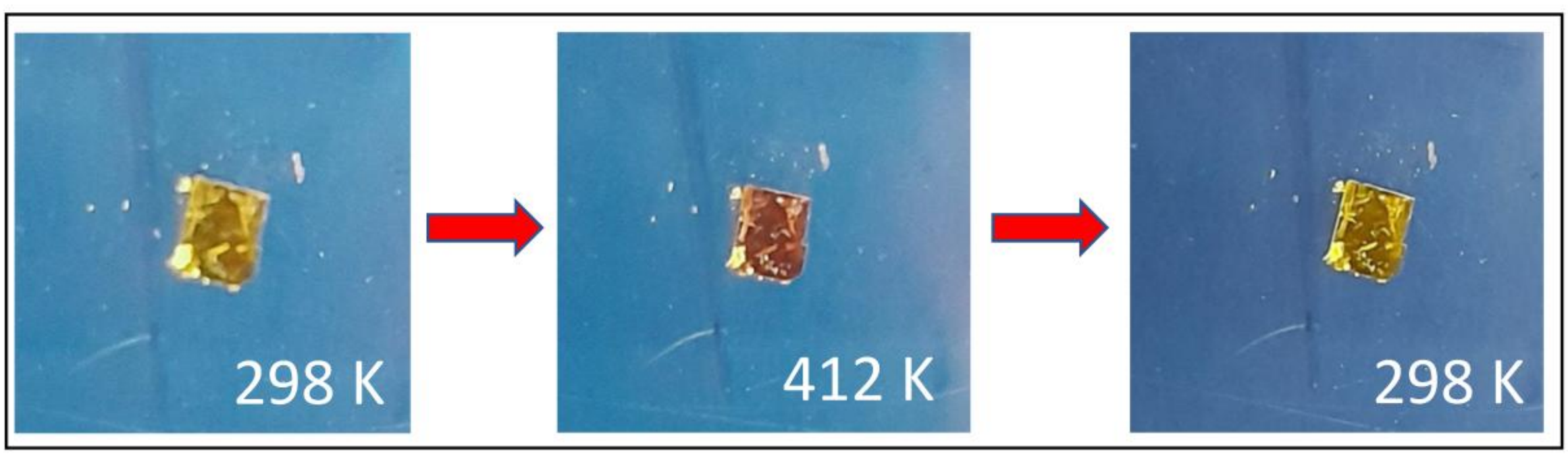

**Figure S5** Thermochromic behaviour of the as grown crystal studied through the temperature of 298K to 413K.

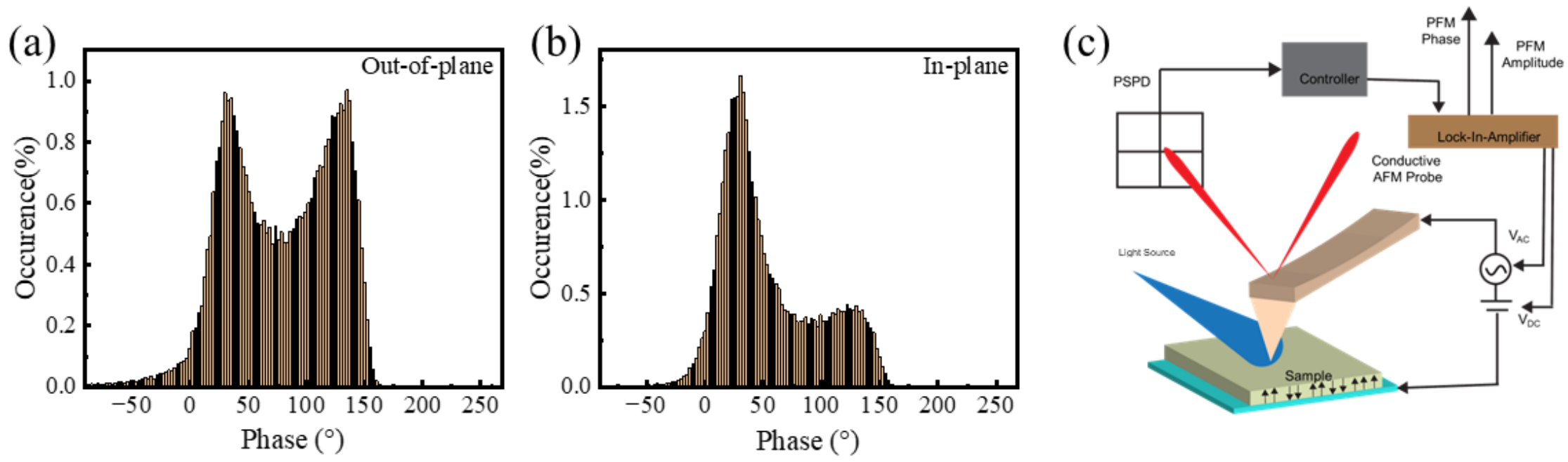

**Figure S6** Population distribution of two different polar domains observed in $(3\text{-}FBA)_2CuCl_4$ (a) out-of-plane and (b) in-plane PFM phase response. (c) Schematic diagram of PFM